\documentclass[12pt]{article}
\usepackage{blindtext}
\title{Turbulence Closure modeling with machine learning approaches: A Perspective}
\author{Sharath S. Girimaji\\ Ocean Engineering Department\\ Texas A \& M University}

\date{\today}

\begin{document}
\maketitle

%\section{Abstract}
\begin{abstract}
  
Turbulence closure modeling using machine learning is at an early crossroads. The extraordinary success of machine learning (ML) in a variety of challenging fields has given rise to justifiable optimism regarding similar transformative advances in the area of turbulence closure modeling. However, by most accounts, the current rate of progress toward accurate and predictive ML-RANS (Reynolds Averaged Navier-Stokes) closure models has been much slower than initially projected.
The slower than expected rate of progress can be attributed to two reasons: high initial expectations without a complete comprehension of the complexity of the turbulence phenomenon, and the use of ML techniques in a manner that may be inconsistent with turbulence physics.
To do full justice to the potential of data-driven approaches in turbulence modeling, this article seeks to identify the foundational physics challenges underlying turbulence closure modeling and assess whether machine learning techniques can effectively tackle them. Drawing analogies with statistical mechanics and stochastic systems,  the key physical phenomena and mathematical limitations that render turbulence closure modeling complicated are first identified. The inherent capability of ML to overcome each of the identified challenges is then investigated.
The conclusions drawn highlight the limitations of ML-based closures and pave the way for a more judicious framework for the use of ML in turbulence modeling. As ML methods evolve (which is happening at a rapid pace) and our understanding of the turbulence phenomenon improves, the inferences expressed here should be suitably modified.

%This article seeks to explicate the fundamental challenges posed by the turbulence closure problem and examine the ability of ML approaches to address them. By drawing analogy with statistical mechanics, in general, and kinetic theory of gases, in particular, 
%the key physical phenomena and mathematical limitations that render turbulence closure modeling intractable are first identified. 
%The inherent capability of ML to overcome each of the identified challenges is then investigated. 
%The inferences point to the limitations of ML-based closures and could lead to a framework for more judicious use of ML for turbulence modeling. As the ML methods evolve, which is happening at a rapid pace, and our grasp of the turbulence phenomenon improves, the inferences expressed here should be suitably modified.
\end{abstract}
%Mainstream research on ML turbulence modeling is less than a decade old, but what is lacking in duration has been amply compensated by the intensity of effort. There is now an adequate body of work in the area for assessing progress and developing a more considered second opinion on the potential of ML to advance the field of turbulence closure modeling.
%This article aims to elucidate the fundamental challenges posed by the turbulence closure problem and assess the capability of ML approaches to address them. 

\section{Introduction}

The phenomenon of fluid turbulence is completely described by the Navier-Stokes equations in the continuum flow regime. Non-linearity and non-locality are two defining characteristics of the Navier-Stokes equations. Direct numerical simulation (DNS) of high Reynolds number flows is infeasible due to the exceptionally large degrees of freedom (scales of motion) engendered by non-linearity. Interestingly, non-linearity is not the main reason for the intractability of the Navier-Stokes equation. The three-dimensional Burgers equation which contains the same type of non-linear term permits an analytical solution \cite{GAO2017255}. Rather, it is the non-locality of the pressure effect which is governed by the elliptic Poisson equation in the incompressible flow limit that renders analytical treatment of the Navier-Stokes equation extremely complicated. The non-linearity compounds the non-locality effects leading to a chaotic system that has earned turbulence the reputation of being the last unsolved problem in classical physics. Due to this complexity, the development of adequately accurate reduced-order models has proven challenging. 

Turbulence closure modeling, which is a form of reduced-order modeling, begins with statistically averaging or filtering the Navier-Stokes equation.  Filtering the Navier-Stokes equations reduces the degrees of freedom (or scales of motion) at the cost of introducing unclosed higher-order statistics.  The complexity of the physics incumbent in the unclosed statistics escalates with increasing filter width or decreasing degrees of freedom.  In the commonly used single-point RANS (Reynolds-averaged Navier-Stokes) approach, all of the fluctuations of the flow field are eliminated using a suitable averaging procedure. Modeled equations are solved only for the mean flow field variables. The main advantage of the RANS approach is its low computational cost. However, the physics embedded in the various unclosed terms in the evolution equation can be very intricate leading to modeling difficulties. At the other end of the modeling spectrum, the filter width in LES (large eddy simulations) is significantly narrower. Thus, LES is computationally more expensive, but the physics resident in the various unclosed terms of the equation is more straightforward. Well-resolved LES is still computationally prohibitive for engineering design computations. Thus, despite inherent shortcomings, turbulence modeling at the RANS level remains a subject of much contemporary interest due to its practical utility. The various aspects contributing to the complexity of RANS modeling will be discussed in detail later in the paper.

The remarkable success of machine learning in diverse fields such as pattern identification, speech recognition, automation, and finance,  witnessed over the last decade, has raised expectations for similar advancements in the area of turbulence closure modeling. In recent years, a substantial volume of research in this area has been reported in the literature. Supervised machine learning, particularly with deep neural networks representing constitutive equations, has gained popularity as a potential means of enhancing the accuracy of RANS closure models.
However, the existing body of literature points to mixed success at best.
While ML-RANS models have shown some improvement, they are still far from accurately predicting key characteristics of complex, unseen flows.  At a recent conference, it was concluded that 'the field of machine learning has a long way to go before general improvements to turbulence models may be seen. In other words, the payoff, if any, will probably not be immediate' (NASA Symposium on Turbulence Modeling: Roadblocks, and the Potential for Machine Learning -- NASA/TM-20220015595. Authors: Rumsey and Coleman. Date of publication: November 2022).

We propose that two principal reasons for the underperformance are unrealistic expectations without a full understanding of the nature of turbulence modeling complexity and the utilization of machine learning tools in a manner inconsistent with flow physics, even when success may be possible. In order to derive the full benefits of machine learning for turbulence closure modeling applications, it is vital to have a broad perspective and a comprehensive understanding of the foundational challenges. Proceeding to closure modeling without such comprehension can lead to unsatisfactory outcomes and ultimately be a disservice to both the machine learning and turbulence modeling fields.

%The remarkable success of machine learning in many fields (such as pattern identification and speech recognition) witnessed over the last decade has led to expectations of similar advancement  in the area of turbulence closure modeling. Over the last few years, a large volume of research in this area has been reported in literature. Supervised machine-learning with deep neural networks for representing constitutive equations have gained popularity for enhancing RANS computations.  While the ML-RANS models have shown some improvement, they are still far from being able to accurately predict key characteristics of complex unseen flows. The body of literature thus far points to mixed success at best. At a recent conference, it was concluded that {\em `the field of ML has a long way to go before general improvements to turbulence models may be seen. In other words, the payoff (if any) will probably not be immediate'} (NASA Symposium on Turbulence Modeling: Roadblocks, and the Potential for Machine Learning --  NASA/TM-20220015595. {\em Authors} - Rumsey and Coleman. {\em Date of publication} - November, 2022). We propose that two of the principal reasons for the under performance are: unrealistic expectations without full cognizance of the nature of turbulence modeling complexity; and, utilization of the ML tools in a manner that is inconsistent with flow physics even when success may be  possible.
%It is unclear if the slow slow rate of progress is due to inherent limitations of ML techniques, or sub-optimal use of these methods at the current stage of development. 

The objective of this article is to identify the intractable features of turbulence closure physics by drawing an analogy with statistical mechanics and stochastic systems. Then, 
 we seek to examine the ability of ML-based closures to adequately address the challenging features.  This is accomplished by seeking answers to five key questions:
\begin{enumerate}
\item When and why do traditional RANS closures fail?
\item How is turbulence different from other problems where ML has enjoyed success?
\item What are the challenges to developing generalizable ML turbulence closure models?
\item What are the data requirements for developing predictive ML closure models? Is it feasible to acquire the required data?
\item What key lessons have been learned thus far, and what are some meaningful directions for future research? 
\end{enumerate}
It is important to note that the paper is not about machine learning techniques. Rather,  it addresses foundational flow physics issues pertaining to the use of ML techniques for developing effective RANS turbulence closures. The paper seeks to explicate the challenges associated with using neural networks for constitutive relations and other closure terms. The insight developed herein can be useful for a broader range of ML applications in the field of turbulence modeling.

In Section 2 of the paper, we provide a brief discussion of the full governing equations and present the unclosed reduced-order equations. In Section 3, we explain the challenges of the turbulence closure problem in the context of statistical mechanics and stochastic processes to better comprehend the degree of closure modeling difficulty. Section 4 examines the capabilities and shortcomings of traditional closure models in addressing the complex features of turbulence. With the background and context established, Section 5 seeks answers to the five ML-related questions listed above. We conclude in Section 6 with a brief summary. Readers may find it useful to peruse the Conclusions section first before going through the remainder of the paper.

% In Section 2 of the paper, we present a brief discussion of the features that render  turbulence closure modeling a complex problem. In Section 3, the challenges of the turbulence closure problem are explained in the context of statistical mechanics (kinetic theory of gases) to better comprehend the degree of intractability. The capabilities and short-comings of traditional closure models to address the complex turbulence features are examined in Section 4. Having established the background and context, the answers to the five ML-related questions listed above are sought in Section 5. We conclude in Section 6 with a brief summary.

\section{Governing equations and closure modeling}

The Navier-Stokes equation is the statement of momentum balance in fluid flow and is given by:
\begin{equation}
\label{INS}
\frac{\partial \rho U_i}{\partial t} + U_j \frac{\partial \rho U_i}{\partial x_j} = -\frac{\partial p}{\partial x_i} +
\frac{\partial}{\partial x_j} \left(\mu \frac{\partial U_i}{\partial x_j}\right),
\end{equation}
where $\rho$ is the fluid density, $U_i$ is the velocity field, $p$ is the pressure field, and $\mu$ represents fluid viscosity. The rate of change of inertia per unit volume (advective term) of a fluid particle on the left-hand side of the equation is balanced by the sum of pressure and viscous forces on the right side.
In compressible flows, the density and pressure fields are determined from the conservation of mass and energy equations. In incompressible turbulence, the density is uniform (constant), and pressure is determined as a solution of the elliptic Poisson equation given by:
\begin{equation}
\label{IP}
\frac{\partial^2 p}{\partial x_j \partial x_j} = - \rho \frac{\partial U_i}{\partial x_j} \frac{\partial U_j}{\partial x_i}
\end{equation}
The complexity of turbulence arises from the non-linearity of the advection term and the non-locality of the elliptic pressure equation (\ref{IP}). While non-linearity leads to a proliferation of scales, non-locality renders the pressure-velocity interactions very complicated (\cite{mishra_girimaji_2013}, \cite{mishra_girimaji_2014}, \cite{mishra_girimaji_2017}, \cite{PhysRevE.92.053001}, \cite{mishra2010}, \cite{mishra2016}). In combination, these phenomena can lead to broadband chaotic behavior of the velocity field. The existence or smoothness of solutions in general three-dimensional flows cannot be formally proven.
The range of scales generated due to non-linear phenomena is dependent on the Reynolds number, which is the ratio between inertial and viscous effects:
\begin{equation}
\label{re}
Re \equiv \frac{\rho U L}{\mu},
\end{equation}
where $U$ and $L$ are the characteristic velocity and length scales. With increasing Reynolds number, the range of scales increases. At high enough Reynolds numbers, the small scales are well separated from the large scales. The large scales of motion are strongly dependent on the flow geometry.
On the other hand, as hypothesized by Kolmogorov (\cite{1941DoSSR..32...16K}, \cite{kolmogorov_1962}), the small scales exhibit nearly universal statistical behavior in most high Reynolds number flows of practical interest. 

\paragraph{Coherent structures.} 
The flow-geometry-dependent largescale features can take the form of organized patterns called {\em coherent structures}. Many flows of practical interest, in engineering and nature, exhibit an intriguing mix of coherent structures and incoherent stochastic flow features.  The distinction between the coherent and stochastic fields is important as they have completely different effects on the mean flow. Due to its incoherence, the stochastic turbulence field is `dynamically passive'.  Its overall effect on the mean flow field can be characterized as diffusive and dissipative. From a thermodynamic viewpoint, the action of the stochastic part is irreversible and thus lends itself to easier closure modeling. In contrast, the coherent part can be 'dynamically active'. The effect of coherent structures is generally one of advective stirring of the mean flow rather than diffusion. In a statistical sense, the stirring action is reversible and can transfer energy from small to large scales of motion. 

The coherent structures typically arise from largescale instabilities and their flow physics is strongly influenced by multi-point space and time correlations. For a given flow geometry, the coherent structures can also change significantly with the Reynolds number. Even the simple flow past a circular cylinder exhibits different types of coherent structures depending upon the Reynolds number \cite{doi:10.1146/annurev.fl.28.010196.002401}.  Thus, the coherent velocity field is not only geometry-dependent but can also exhibit strong bifurcations as a function of Reynolds number.

Depending upon the flow type, the stochastic component of the flow field can be present in any scale of fluid motion,  but the coherent structures are typically resident in large scales. Girimaji and Zhou \cite{girimaji1996analysis} provide a detailed discussion on the size of scales that have a stirring (dispersive) effect and those that are likely to be purely diffusive. In nearly all homogeneous flows and simple inhomogeneous flows, there are no large coherent structures, and the fluctuations at all scales can be treated reasonably as stochastic.
Even in practical flows with coherent structures, the small scales generated by the non-linear cascade process can be treated as stochastic.

In many engineering applications, the flows can be further complicated by the effects of compressibility, combustion, added body forces, streamline curvature, and system rotation. For the sake of simplicity, we restrict our discussion to incompressible turbulent flows without added complexities.  
Even incompressible flows of engineering relevance are generally too expensive to compute directly from first principles (direct numerical simulations, DNS) due to the broad spectrum of scales. The different scales of motion can be viewed as distinct degrees of freedom. For practical flow computations, reduced-order models that entail the elimination of a range of scales (by averaging or filtering) are required. 
In the past few decades, researchers have primarily followed a hypothesis-driven 'traditional' closure modeling approach. As highlighted in the Introduction, more recently, data-driven techniques have been employed to develop reduced-order models. The objective of this study is to conduct a comparative assessment of these two approaches.

\vspace{.1in}

\subsection{Closure modeling}
To effectively compare the traditional and machine-learning-based approaches, it is vital to begin with a comprehensive understanding of the model development process.
The  closure model development procedure  in both approaches involves three key steps: 
\begin{enumerate}
    \item {\bf Order reduction by filtering or averaging}. In this first step, select scales of motion are eliminated from the governing equations to reduce the computational burden. Due to the nonlinearity of the equations, unclosed higher-order statistics appear in the reduced-order description. 
    \item {\bf Mathematical form of unclosed statistics}.  In the next step,  mathematical expressions are proposed to model higher-order statistics to close the equations describing the reduced-order system. 
    \item {\bf Determining coefficients.} The final step entails determining the coefficients in the closure model expressions to ensure the broad applicability of the reduced-order models. 
\end{enumerate} 
It is important to recognize that each of these steps introduces approximations and errors. This groundwork sets the stage for a thorough comparison between traditional and machine-learning approaches in the remainder of the paper.

Order reduction can be carried out at different degrees of scale elimination. Here we present the model equations at the two extreme levels of order reduction, namely RANS and LES. The terms in the RANS/LES equations requiring closure modeling are then identified. The primary emphasis of this article revolves around one-point closures, while relevant discussions on key features of two-point closures will also be included as appropriate.

\subsubsection{RANS approach:} 

The Reynolds averaged Navier-Stokes (RANS) method represents the most commonly used approach in practical flow computations. In RANS, all of the fluctuating scales of motion are eliminated by averaging.
It is  expedient to decompose the velocity field into mean and fluctuating parts:
\begin{equation}
\label{decomp}
    U_i \equiv \overline{U_i} + u_i; \;\; 
%    \mbox{where} \; \:  u_i \equiv  V_i + v_i.
\end{equation}
Here, $U_i$ is the instantaneous velocity; $\overline{U_i}$ is the Reynolds-averaged velocity field and $u_i$ is the fluctuating field. It is important to recognize that $u_i$ can include contributions from the coherent structures. A similar decomposition can be performed for other flow variables. For compressible flows, density-weighted (Favre) averaging is more appropriate. Here, we will restrict our focus to incompressible flows.
The transport equation for the mean flow field is derived by suitably averaging the Navier-Stokes equations - 
\begin{equation}
\label{RANS}
    \frac{ \partial  \rho \overline{U_i}}{\partial t} + \overline{U_j} \frac{\partial 
    \rho \overline{U_i}}{\partial x_j} = -\frac{\partial \overline{p}}{\partial x_i} - \frac{\partial \rho R_{ij}}{\partial x_i} 
   +  \frac{\partial}{\partial x_j} (\mu \frac{\partial \overline{U_i}}{\partial x_j} ),
\end{equation}
The effect of the fluctuating velocity field manifests on the mean-field {\em via}  the Reynolds stress tensor $R_{ij}$:
\begin{equation}
    \label{Rij}
    R_{ij} \equiv \overline{u_i u_j}. 
\end{equation} 
The Reynolds stress can be decomposed into a scalar measure, turbulent kinetic energy $K$; and, the directional anisotropy tensor $b_{ij}$: 
\begin{equation}
    \label{bij}
    b_{ij} \equiv \frac{\overline{u_i u_j}}{2K} - \frac{1}{3} K \delta_{ij};  \;\;\mbox{} \;\; K = \frac{1}{2} \overline{u_i u_i}.
\end{equation}
%RANS computation entails numerically solving the mean-flow evolution equations within the flow domain of interest. 
The  Reynolds stress ($R_{ij}$)  evolution equation (RSEE) can be derived from the instantaneous flow field equation (\ref{INS}) and RANS  equation (\ref{RANS}) :
\begin{equation}
    \label{RSEE}
   \frac{ \partial \rho R_{ij}}{\partial t} + \overline{U_k} \frac{\partial \rho  R_{ij}}{\partial x_k} = 
       P_{ij} + \Pi_{ij} - \varepsilon_{ij} + \frac{\partial T_{ijk}}{\partial x_k}.
\end{equation}
In the RSEE, the various terms on the right side are the production ($P_{ij}$), pressure-strain correlation ($\Pi_{ij}$), dissipation ($\varepsilon_{ij}$), and turbulent transport ($T_{ijk}$).

In the RANS approach,  closure modeling can be undertaken directly for $R_{ij}$ or by solving its evolution equation. If $R_{ij}$ is modeled directly, then a constitutive relation between stress and the mean-velocity field must be developed. Additional modeled equations for turbulent velocity and length/time scales may be needed. If, on the other hand, the RSEE is solved to obtain $R_{ij}$, closure models for pressure-strain correlation, dissipation, and turbulent transport are needed. The latter approach is generally referred to as Second Moment Closure (SMC).
Importantly, the RANS method entails solving a dynamical system of equations, wherein each equation contains multiple unclosed terms requiring modeling. Two classes of closure models and coefficients must be recognized: the constitutive closure coefficients (CCC) in the Reynolds stress and pressure-strain correlation closures; and, the transport closure coefficients (TCC) in the modeled transport equations of dissipation and kinetic energy. 

\subsubsection{LES approach:} In the large eddy simulation approach, only the smallest scales of motion are eliminated by filtering.  In this case, the total velocity field is decomposed as:
\begin{equation}
    \label{res}
    U_i = \tilde{U_i} + w_i,
\end{equation}
where tilde ($\tilde{.}$) represents the filter operator,  $\tilde{U_i}$ is the largescale resolved field and $w_i$ is the velocity field of the unresolved scales of motion. In this case, the coherent structures are resident in the resolved field.
The equations for the resolved field can be derived formally by filtering the Navier Stokes equations (\cite{deardorff1970numerical}, \cite{germano_1992}). For the case of a spatially and temporally invariant filter, the resolved flow equation can be written as \cite{germano_1992}:
\begin{equation}
    \label{Germano}
    \frac{ \partial  \rho \tilde{U_i}}{\partial t} +  \frac{\partial 
    \rho \tilde{U_i} \tilde{U_j}}{\partial x_j} = -\frac{\partial \tilde{p}}{\partial x_i} - \frac{\partial \rho \tau_{ij}}{\partial x_i} 
   +  \frac{\partial}{\partial x_j} (\mu \frac{\partial \tilde{U_i}}{\partial x_j} ).
\end{equation}
The unresolved ($w_i$) small scales influence the resolved velocity field evolution through the subgrid (SGS)  or sub-filter stress (SFS) $\tau_{ij}$:
\begin{equation}
    \label{SGS}
    \tau_{ij} \equiv \widetilde{U_i U_j} - \tilde{U_i} \tilde{U_j}.
\end{equation}
 The most striking feature of the resolved flow equation (\ref{Germano}) is that it is similar in form to that of the RANS equation (\ref{RANS}). Indeed, the evolution equation for the subgrid stress $\tau_{ij}$ can be cast in a form similar to RSEE \cite{germano_1992}. 
 However, the magnitude of $\tau_{ij}$ and the physics incumbent in the various terms in the $\tau_{ij}$ evolution equation depend upon the size of the filter width.
In the limit of large filter width (averaging), all fluctuating scales are eliminated from consideration. Then, $\tau_{ij}$  tends to $R_{ij}$ and the resolved flow equation asymptotes to the RANS equation (\ref{RANS}). 
%Thus, averaging can be considered as an extreme limit of filtering.
In the opposite limit of very small filter-width, $\tau_{ij}$ vanishes, and equation (\ref{SGS}) reduces to the instantaneous Navier-Stokes equation (\ref{INS}). 

As in the RANS approach, SGS stress $\tau_{ij}$ can be modeled directly or by solving its evolution equation. For small filter widths,
the sub-grid stress for LES can be closed with a simple algebraic model as most of the flow complexity is incumbent in the resolved scales.
The so-called scale-resolving simulations (SRS) represent an intermediate level of filtering between RANS and LES. The degree of difficulty of the SRS closure model depends upon the implied filter width. When the filter-width is large, a RANS-type closure model that entails solving a dynamical system of equations becomes necessary. Thus, the challenge of closure modeling is influenced by both the intricacies of the flow and the cut-off length scale. 

\paragraph{Turbulence Complexity.}
A key obstacle in developing broadly applicable turbulence models stems from the non-universality of large scales, especially in the presence of flow-dependent coherent structures. Therefore, we propose a turbulence modeling complexity metric based on the flow-dependent coherent structure content within the velocity field.
First, the velocity field is notionally decomposed as follows \cite{doi:10.1063/1.866322}:
\begin{equation}
\label{triple} 
     u_i \equiv  V_i + v_i
\end{equation}
where the coherent part is given by $V_i$ and the stochastic part is denoted by $v_i$.
The contributions of the coherent and stochastic flow fields to the Reynolds stress can be partitioned:
\begin{equation}
\label{RSC}
 R_{ij} = \overline{u_i u_j} = \overline{V_i V_i} +  2 \overline{V_i v_i} + \overline{v_i v_i}.
\end{equation}
%Each term on the right side of the equation (\ref{RSC}) presents different degree of closure challenge. 
%In most flows of practical interest, the dynamically active coherent structures reside exclusively in the largescales. 
%The cross-covariance term $\overline{V_i v_i}$ is particularly difficult to analyze. 
If indeed the coherent structures are resident entirely in the large scales and the stochastic part in the small scales, then the cross-variance term will be small due to length scale  mismatch:
\begin{eqnarray}
    \overline{V_i v_i} \sim 0.
\end{eqnarray}
%While the above simplification may not be necessarily valid, it permits easier qualitative analysis.
The relative contributions of coherent structures and stochastic features can be a strong function of the flow type. As mentioned earlier, flows in which the coherent part of the field dictates the overall dynamics can be classified as complex. 
We propose the fraction of coherent field contribution to the total turbulent kinetic energy ($f_c$) as a key metric for quantifying the degree of closure modeling difficulty:
\begin{equation}
\label{cindex}
    f_c \equiv \frac{\overline {V_i V_i}}{\overline{u_i u_i}} \sim 
    \frac{\overline{V_i V_i}}{\overline{V_i V_i} + \overline{v_i v_i}}.
\end{equation}
The lower the value of $f_c$, the lesser is the importance of coherent structures and long-range interactions. Clearly, $f_c$ will be small and spatially uniform in homogeneous turbulent flows.
In complex flows, $f_c$ can be large and vary in space depending upon the location of the coherent structures.   Furthermore, the level of complexity will be scale-dependent. A flow with large unsteady coherent structures could be complex in the large scales and still exhibit stochastic behavior at the small scales. It is important to note that the above complexity characterization is not rigorous. Nonetheless, such qualitative distinction is of much value for the development of closure models.

\paragraph{Elements of Closure Modeling.}
Once the reduced-order equations are developed, successful closure modeling of the unclosed terms depends on many factors. We highlight four key required elements here. 
\begin{enumerate}
    \item {\bf Comprehension of physics.} A deep understanding of the physics underlying the various terms in the averaged/filtered equations is crucial.  It is also vital to identify quintessential mathematical and physical elements that must be conveyed to the closure models.
    \item {\bf Leveraging accumulated data and assimilated knowledge.} Over the last several decades, much turbulence data has been accumulated from experiments and high-fidelity simulations. Useful hypotheses and laws that shed light on turbulence beyond what is known directly from equations have been developed. Leveraging this knowledge in the model development process is very valuable.
    \item {\bf Abstraction of general principles.}  Utilizing the knowledge assimilated from the first two elements, it is important to generate a higher level of understanding to serve as closure modeling principles or guidelines that can lead to generalizability. This requires critical and abstract analysis of the equations, data, and hypotheses.
    \item {\bf Effective and innovative mathematical tools.} To incorporate identified features into the closure model foundation, effective tools must be employed. Given the complexity of turbulence, innovative approaches may be necessary.
\end{enumerate}
In the next section, we will explore the specific physical and mathematical features that render the turbulence closure model development challenging. Then, we will proceed to analyze how traditional and ML methods address the challenges.

\section{Closure modeling challenges}

The challenges posed by the RANS equations are distinctly different from those of the instantaneous Navier-Stokes equations (DNS). The RANS equations lack both the broad range of scales and the chaotic tendencies observed in DNS. However, the flow physics incumbent in the unclosed RANS/LES terms is significantly more complicated than any term in the original Navier-Stokes equation. Therefore, averaging shifts the challenge from the large computational effort required for solving instantaneous Navier-Stokes equations to modeling intractable flow physics in the unclosed terms of the RANS equations. Turbulence modeling is not the only problem in classical physics that encounters such a closure issue. Indeed, the field of statistical mechanics, in general, and kinetic theory of gases (KTG), in particular, addresses similar challenges. The discipline of stochastic dynamical systems also shares many features in common with turbulence modeling. To gain a broad perspective on the challenges of turbulence closure modeling, it is useful to derive insights from these two fields of classical physics.

%The root causes of complexity of the RANS equations are distinctly different from those of the instantaneous Navier-Stokes equations (DNS). The RANS equations exhibit neither the broad range of scales nor the chaotic tendencies. As a consequence of averaging, the  flow physics represented by the unclosed terms becomes substantially more complicated.  Thus, averaging shifts the burden from large computational effort required for solving instantaneous Navier Stokes equations to accurate representation of the flow physics of the unclosed terms of RANS. Turbulence modeling is not the only problem in classical physics that encounters such a closure problem.  Indeed, the field of statistical mechanics, in general, and kinetic theory of gases (KTG), in particular, encounters similar challenges. The field of stochastic dynamical systems also shares many features in common with turbulence modeling. In order to gain a broad perspective of the turbulence closure modeling challenges, it is useful to gain insights from these two classical fields of physics.

\subsection{Insights from statistical mechanics}
Statistical Mechanics is a formal framework whose objective is to determine the macroscopic (ensemble or aggregate)  behavior of systems that are composed of a large number of interacting microscopic elements.  The physical laws governing the behavior of microscopic elements are taken to be known.  Then, statistical tools and probability theory are used to determine the governing equations at the macroscopic level. The KTG is one of the earliest examples of a statistical mechanics approach for scaling up a system description from microscopic-molecular to macroscopic-continuum scales. 
Turbulence modeling can also be regarded as a statistical mechanics approach that seeks to describe mean-flow (RANS) or largescale (SRS) behavior of a turbulence field commencing from the physical laws that govern the full flow evolution (i.e, the Navier-Stokes equation).

%KTG can yield valuable insight into turbulence modeling challenges for practical flow applications. 

Molecular dynamics (MD) describes the behavior of a gaseous system of atoms and molecules at the microscopic level.  MD numerical simulations of gaseous systems are not only expensive, but also the detailed description of molecular motion generated by such computations are not required for most applications.  This is analogous to DNS of turbulence which is expensive \textit{and} the detailed description of the velocity field that it generates is unnecessary for most applications.  Statistical mechanics approach to gaseous system then seeks a reduced order description. Central to order reduction is the Boltzmann equation which is derived from the physical laws at the microscopic level subject to two main simplifications -  the `dilute gas' and  `molecular chaos'  assumptions.  The dilute-gas simplification precludes the simultaneous collision of multiple particles. The molecular chaos ansatz guarantees that the velocities of two colliding particles are uncorrelated. In other words, the molecular motion is random and not `coherent'.
   %If the molecular motion is organized, further simplifications are not straight-forward.  

%In principle, the  Boltzmann equation is valid at all degrees of rarefaction and non-equilibrium particle distribution.  Further simplifications are possible in systems that satisfy low Knudsen number ($Kn$) and near-equilibrium conditions.   In low $Kn$ systems,  the microscopic and macroscopic length and time scales are well separated. Equilibrium refers to a system in which the single particle velocity distribution function (vdf) is close to a Maxwellian or Gaussian.  Using Chapman-Enskog analysis, the transport properties of near-equilibrium continuum gases can be derived \cite{chapman1990mathematical}.  

Subject to the above assumptions, the Boltzmann equation is valid at all degrees of rarefaction and non-equilibrium particle distribution. Further simplifications are possible in systems that satisfy low Knudsen number ($Kn$) and near-equilibrium conditions. At low $Kn$ systems, the microscopic and macroscopic length and time scales are well separated. Equilibrium refers to a system in which the single particle velocity distribution function (vdf) is close to a Maxwellian or Gaussian. Subjecting the gas system to these further simplifications and employing the Chapman-Enskog analysis, the transport properties of near-equilibrium continuum gases can be derived
\cite{chapman1990mathematical}.
This leads to conservation equations at the macroscopic level with constitutive relations for the transport of mass, momentum, and energy transport.  At the most elementary level (the highest degree of order reduction), the continuum-scale momentum and thermal balance statements are given by the Navier-Stokes-Fourier (NS-F) equations. Gaseous systems that are slightly or even moderately removed from equilibrium can also be modeled with continuum concepts and the constitutive relations can be derived using extended thermodynamic principles.

% Much like turbulence modeling, the goal of kinetic theory is to
%describe aggregate physical properties or statistics of the system as a function of the relevant field variables at different levels of order reduction. 
% In kinetic theory of gases, a system at or close to equilibrium is  most easily amenable to significant order reduction and straight-forward closure modeling. 
%Using powerful concepts such as `molecular chaos', the evolution of a multi-particle velocity distribution function (vdf) can  simplified to a single-particle vdf leading to the Boltzmann equation. After further assumptions are invoked, Chapman-Enskog analysis is employed to derive the transport properties of gases  The  continuum NS-F equations are analogous to RANS closure models for the mean velocity field in a turbulent flow.
 %When there is multi-particle coherence, the molecular chaos ansatz is not valid. 
% As in turbulence closure equations, the presence of coherence in the field can lead to complexities in reduced order modeling.

If the molecular chaos statement is invalid, indicating coherent motion in microscales, many of the simplifications in the kinetic theory of gases (KTG) cannot be justified. This will introduce complexities in the macroscale description of system behavior \cite{doi:10.1063/1.2916049}. Gas flows that exhibit molecular coherence or non-equilibrium effects often require more intricate theories, involving additional equations \cite{doi:10.1098/rsta.2020.0066}. Therefore, kinetic systems that are far from equilibrium (rarefied flows) or those with long-range interactions (memory effects) may not be amenable to simple constitutive equations. In summary, KTG provides a formal framework for developing reduced-order models, including constitutive relations, for gaseous systems subject to the following conditions: (i) the 'molecular chaos' ansatz; (ii) adequate separation of micro and macro time and length scales of motion; and (iii) the 'near-equilibrium' assumption.

 %We now summarize the various factors contributing toward the complexity and intractability of turbulence closure modeling. To begin with, non-linearity and non-locality are two defining characteristics of the Navier-Stokes equations that lead to a high-dimensional chaotic dynamical system. Despite this complexity, the Navier-Stokes system is Markovian or memory-less in nature. To reduce the order of the system to manageable levels,  filtering/averaging operation is performed. This reduction leads to two important repercussions - the closure problem and non-Markovian memory (NMM) effects. The closure problem is engendered by the non-linearity of the fundamental equations and leads to more unknown variables (statistics) than equations at any level of statistical description.  The spatio-temporal information contained in the eliminated scales manifests as the NMM effect.
% The degree of NMM effects depend upon the type of turbulence. 
% The non-locality of pressure can lead to spatial NMM effects in pressure-related statistics. Departure from equilibrium and coherence in the eliminated field can impart temporal and spatial NMM effects in all unclosed statistics. Thus, turbulence closure modeling must account for the spatio-temporal memory incumbent in all statistics.

The similarities between turbulence closure modeling and KTG are evident from the preceding discussions. Much like KTG, one-point turbulence models aim to describe turbulence at the aggregate statistical level. 
%using simple models to represent the physical manifestations of unresolved scales of motion on the statistics.
Drawing insights from KTG, it can be inferred that turbulence physics may be amenable to straightforward statistical closure modeling under the following conditions: (i) an adequate size separation exists between large and small scales of motion—equivalent to a low Knudsen number; (ii) the phenomenon is reasonably close to an equilibrium state, and (iii) in the absence of coherent flow structures—similar to the molecular chaos ansatz. We will now delve deeper into each of these challenges.

%The similarities between turbulence closure modeling and KTG are evident from the preceding discussions.  Much like KTG, one-point turbulence models seek to describe turbulence at the aggregate statistical level with  simple models to represent the physical manifestations of unresolved scales of motion on the statistics. 
%From KTG, it can be inferred that the turbulence physics may be amenable to {\em straight-forward} statistical closure modeling under the following conditions: (i) adequate size separation exists between large and small scales of motion - equivalent of low Knudsen number; (ii) the phenomenon is reasonably close to equilibrium state; and  (iii) in the absence of coherent flow structures - similar to molecular chaos ansatz.  We will now delve deeper into each of the three aspects. 

\paragraph{Inherent limitations of averaging - Irreversibility.} 
From the outset, it is crucial to acknowledge the inherent limitations of statistical mechanics, arising from considerations of reversibility and recurrence \cite{sep-statphys-statmech, poincare1890probleme, andrews1965statistical}. While microscopic systems exhibit time symmetry (time reversibility), the macroscopic depiction is predominantly irreversible, aligning with the arrow of time dictated by entropy \cite{mackey2011time}. In the context of turbulence modeling, irreversibility resulting from filtering or averaging the Navier-Stokes equations introduces key physical limitations. The excluded scales of motion can exhibit a reversible advective action on the resolved field, whereas statistical models predominantly represent an irreversible diffusive effect. This irreversibility limitation can lead to significant inaccuracies in flows characterized by large coherent structures. The coherent large scales 'stir' the field, whereas the statistical models are constrained to diffuse the field. This limitation persists even with the most advanced closure models \cite{2020APS..DFDS01032G}.

\paragraph{Scale separation.} As discussed in the Introduction, the spectrum of scales observed in a turbulent flow field is contingent upon the Reynolds number. At lower Reynolds numbers, this spectrum is relatively narrow. The spectrum gradually broadens as the Reynolds number increases. %In high Reynolds number turbulence, a distinct separation exists between the large energy-containing scales and the smaller dissipative scales. 
In molecular dynamics, a clear demarcation exists between continuum and molecular scales, justifying the small Knudsen number simplification. However, turbulence presents a continuous and broad spectrum without a noticeable gap between resolved and excluded scales. This lack of a spectral gap precludes a rigorous Chapman-Enskog type of analysis in turbulence. While an approximate analysis is feasible by neglecting the influence of intermediate scales, such a treatment will be valid only at high Reynolds numbers. Clearly, turbulence closure poses further challenges beyond KTG transport formulations.

%As mentioned in the Introduction, the range of scales exhibited by a turbulence flow field depends upon the Reynolds number. At low Reynolds numbers, the range of scales is rather narrow. With increasing Reynolds number, the range broadens. At high Reynolds numbers, there is clear separation between energy containing large scales and dissipative small scales.  Turbulence exhibits a broad and full spectrum with no real spectral gap between resolved and excluded scales. This is in contrast to molecular dynamics, where there is a clear separation between continuum and molecular scales, which justifies the small Knudsen number simplification. Due to the lack of spectral gap, the turbulence phenomenon does not permit a rigorous Chapman-Enskog type of analysis. An approximate analysis is possible by omitting the influence of the intermediate scales, but this is permissible only at large Reynolds number. Thus, turbulence closure is inherently more complicated that KTG transport formulations.

\paragraph{Non-equilibrium turbulence effects.}
In general, the equilibrium state of a dynamical system occurs when the various contributing processes are in balance. In fully developed turbulent flows, where different physical processes reach equilibrium, the velocity distribution function tends to resemble a Gaussian distribution. Turbulent flows can deviate from equilibrium when processes like inertial-pressure-viscous interactions, linear-nonlinear effects, or large-small scale phenomena are not in balance \cite{2020APS..DFDS01032G}. Various forms of non-equilibrium behavior include: (i) transitioning from an arbitrary initial state towards an equilibrium state, as in rapid distortion theory; (ii) migration from one equilibrium state to another; (iii) being in spatial or temporal proximity to a large-scale instability; and (iv) experiencing unsteady forcing. Thus, in transient or unsteady flows and when close to large instabilities, the distribution function can significantly deviate from a Gaussian shape. Many practical flows exhibit one or more of these non-equilibrium features, making their closure modeling challenging.

%In general, an equilibrium state is observed when the various processes contributing to the evolution are  in  balance.  In fully developed turbulent flows, when various physical processes are in balance, the velocity distribution function is generally found to be close to a Gaussian. A turbulent flow can be out-of-equilibrium when the inertial-pressure-viscous processes, or equivalently linear-nonlinear effects, or large-small scale phenomena are out of balance. 
%Different types of non-equilibrium behavior include: (i) evolution from an arbitrary initial state toward an equilibrium state - as in rapid distortion theory ; (ii) migration from one equilibrium state to another;   (iii)  spatial or temporal proximity to a largescale instability; and (iv) unsteadily forced flow. 
%Thus, in transient/unsteady flows and in the proximity of large instabilities, the distribution function can experience significant departures from Gaussian. Most flows of practical interest exhibit one or more of these non-equilibrium features, rendering their closure modeling difficult. 

%Non-equilibrium conditions can occur when turbulence evolves rapidly from one self-similar state to another. To exhibit non-equilibrium properties, the evolution should be rapid relative to the turbulent time scales, because sufficiently slow increase could produce a quasi-static evolution through equilibrium states. Indeed, the goal of the analysis in Ref. [9] was a two-equation model applicable to such slow spectral evolution.}

\subsection{Insights from Stochastic Dynamical Systems}

While turbulence is governed by the deterministic Navier-Stokes equation, there are instances where conceptualizing it as a stochastic process proves insightful. Within the analytical framework of stochastic dynamical systems, we can effectively explore the intricate interplay between 'random' noise and deterministic linear and non-linear effects. In this paradigm, the chaotic behavior arising from multiple degrees of freedom in turbulence is substituted with noise or a stochastic process. This notion allows for the analysis of the turbulence closure development within the framework of stochastic dynamical systems. Such an approach provides a unique perspective to extract additional insights from this field. In the ensuing discussion, we will delve into two important features: the Markovian property and holistic dynamical system modeling.

%Even though turbulence is governed by the deterministic Navier-Stokes equation, it is sometimes useful to consider the phenomenon as a stochastic process.
%The analytical stochastic dynamical system framework is ideally suited for examining the coupling between `random' noise and deterministic non-linear effects.
%In this perspective, the chaotic behavior arising from multiple degrees of freedom in turbulence is substituted with noise or a stochastic process. Consequently, the system of equations for turbulence closure can be analyzed in this framework. This approach opens up the opportunity to gain additional insights from the field of stochastic dynamical systems. Here, we will consider two important features: Markovian property and fixed-point analysis.

%In such characterization, the chaotic action due to multiple degrees of freedom is replaced by noise or a stochastic process. Then, turbulence closure system of equations can be treated as a stochastic dynamical system.  This allows for additional insights to be obtained from the field of stochastic dynamical systems.

\paragraph{Non-Markovian Memory effects in individual unclosed terms.} 
%As a consequence of averaging or filtering, certain fundamental characteristics of the RANS and LES  equations are profoundly different from those of Navier-Stokes equations  \cite{2020APS..DFDS01032G}, \cite{mishraepistemic}. 
The instantaneous Navier-Stokes equations possess the Markovian property - the future state of the full turbulent flow field depends upon the past only through the present full state. Thus, as in DNS, a knowledge of the present state of the full velocity flow field is sufficient to completely and uniquely determine the state of flow in the next instant. As a consequence of averaging, the RANS/SRS equations are not Markovian --  the evolution of statistical or filtered variables of the flow field can depend upon the history (memory) of previous statistical states of the flow field \cite{LUMLEY1979123}, \cite{2020APS..DFDS01032G}, \cite{mishraepistemic}. 
%Solution of a non-Markovian equation requires knowledge of the memory of the process. 
The flow physics eliminated by averaging/filtering manifests as the non-Markovian memory (NMM)  effects embedded in the unclosed terms. The NMM effects can lead to strong non-local spatio-temporal influence. 
Different statistical spatiotemporal histories can lead to different outcomes of RANS calculations.  {\em It is important to note that the NMM spatio-temporal non-locality effects in reduced-order equations are fundamentally different from the pressure non-locality of the instantaneous Navier-Stokes equation.} The NMM effects arise solely due to the averaging operation, whereas the pressure effect in the instantaneous Navier-Stokes equations is still Markovian in nature. While both non-local effects pose challenges, NMM physics can be critically important in the closure modeling of flows with strong spatio-temporal coherence. The NMM effects manifesting through the pressure correlations contain both types of non-local effects and, hence, can be very challenging \cite{mishra_girimaji_2013, mishra_girimaji_2014}.

The extent of  NMM spatio-temporal non-locality influence depends upon (i) the magnitude of the filter width; and, (ii) the degree of spatio-temporal coherence of the eliminated field \cite{PhysRevE.70.045101}.  
The more coherent the eliminated field, the larger will be the memory effect. Consequently, RANS closure modeling of flows with large coherent structures will be more complicated. 
On the other hand, the smaller the filter width, the narrower the range of eliminated scales, and the weaker the influence of the NMM effect. Thus, the LES closure models are less affected by spatio-temporal non-locality than the RANS equations. 
{\em Overall, it bears reiterating, that the filtering operation reduces the excessive computational requirement of the Navier-Stokes equation and transfers the burden to closure modeling of complex unclosed terms (with NMM effects) in the reduced-order equations.}  

 \paragraph{Holistic dynamical system modeling.}  Kinetic theory entails the development of constitutive equations for the transport of mass, momentum, and energy.  The diffusivity, viscosity, and conductivity are based upon fluid properties such as mean free path (characteristic length scale) and peculiar velocity (characteristic velocity scale). Thus, the transport coefficients are exclusively functions of the fluid properties.
 On the contrary, turbulent transport coefficients are strong functions of the flow rather than the fluid. In this case, the turbulent length and velocity scales must be computed from their respective governing differential equations. Thus, the closure models for constitutive relations and characteristic length/time scales must be developed simultaneously. Accurate modeling of each individual phenomenon in isolation cannot guarantee a favorable outcome of a RANS computation. The interplay among the different closure terms and the relationships between the coefficients (TCC and CCC identified earlier)  must be recognized and captured in the modeling process. The overall quality of RANS  prediction depends upon the holistic behavior of the entire dynamical system of equations.  Thus, turbulence closure modeling must be performed in a manner that not only leads to an acceptable approximation of the individual terms but perhaps more importantly, guarantees reasonable holistic behavior of the entire dynamical system \cite{article}, \cite{girimaji_2000}. 

After outlining some of the fundamental challenges in turbulence modeling, we will now explore the difference between traditional and data-driven approaches to closure modeling.

%To assess the overall predictability of a flow, we propose the following `$Re$ vs. $f_c$' map.
%\begin{figure}[ht!]
%\centering
%\includegraphics[width=90mm]{slide1.jpg}
%\caption{A simple caption \label{overflow}}
%\end{figure}

\section{Traditional turbulence closure modeling}

Turbulence closure modeling, as mentioned earlier,  can be performed at different levels of sophistication. The approaches can be broadly divided into three categories: (i)  RANS one-point closures -- which solve evolution equations for fully-averaged one-point statistics. Two equation models and Second-moment closures (SMC) are prime examples of this approach; (ii) Two-point closures --  which solve evolution equations for fully averaged two-point velocity correlations; and (iii) Filtered Navier-Stokes approaches -- which resolve a part of the flow field and model the remainder. Large eddy simulations (LES) and various scale resolving simulations (SRS) approaches belong in this third category.

In the RANS approach, only the mean flow field is computed directly. The effect of the fluctuating field on the mean flow manifests {\em via} a modeled Reynolds stress term. The drastic elimination of all scales of fluctuations results in a significant reduction in computational cost at the expense of closure modeling errors. At the other extreme,
LES  entails filtering out only the dissipative scales of motion. This results in significant simplification of the closure modeling simplicity at the expense of high computational cost. Recent years have witnessed the emergence of the SRS methods that seek to strike an optimal compromise between RANS and LES, by directly computing only the dynamically active large scales. The key features of the three approaches are presented in the remainder of the section. The two-point closures also do not explicitly resolve any fluctuating scales of motion but offer a truer description of the fully averaged field by computing scale-dependent statistics. 

\subsection {RANS one-point closures}

We begin with a discussion of the Second Moment Closure (SMC - equation \ref{RSEE}) that requires the computations of seven extra modeled transport equations aside from the mean flow equations. Then we proceed to briefly examine the simpler two-equation RANS methods which postulate an algebraic equation for the constitutive relation and solve transport equations for only the turbulent length and time scales.
%We will now examine the closure challenges in representing the constitutive relations with a single point closure for the four types of the velocity fields described in the previous section.

\subsubsection{Second Moment Closures.}
In the SMC approach, the unclosed terms in the RSEE (equation \ref{RSEE})  are modeled, and the resulting equations for the components of $R_{ij}$ tensor are solved computationally.  The reader is directed to the following reviews for a detailed account of the development of the SMC approach over the decades \cite{reynolds1976annual}, \cite{LUMLEY1979123}, \cite{speziale1991analytical} and  \cite{girimaji_2000}. It is important to recall that all closure terms can potentially include strong NMM effects in flows with coherent structures. 
Here, we briefly describe the development of the various closure models in the SMC approach focusing on the key assumptions.

\paragraph{Turbulent Transport modeling.} The Reynolds stress transport term is typically modeled with gradient-transport assumption. 
The gradient transport closure expression involves spatial derivatives of Reynolds stresses and can be expressed in the most general form as \cite{speziale1991analytical}:
\begin{equation}
    \label{grad-trans}
    T_{ijk} \sim {\cal C}_{ijklmn} \frac{\partial R_{lm}}{\partial x_n}.
\end{equation}
Here, ${\cal C}_{ijklmn}$ is a sixth-order turbulent diffusion tensor which requires further modeling.
This mathematical form represents a drastic simplification of transport physics wherein stirring action is completely precluded and only diffusive effects are represented. While this form does incorporate some spatially non-local physics, it cannot account for most NMM effects in the presence of coherent structures.  In most practical models, the closure of ${\cal C}_{ijklmn}$ is significantly simplified using symmetry and isotropy arguments. Further details can be found in \cite{1975JFM....68..537L} and \cite{LUMLEY1979123}. 
%It is important to note that these gradient diffusion models can account only for diffusive effects and not for dispersive flow physics.

\paragraph{Dissipation modeling.} It is possible to derive an exact evolution equation for the dissipation tensor from first principles. However, the resulting equation is too complicated to be practically useful for closure modeling. Instead, a much simpler approach is used in practical flow modeling. First, the Kolmogorov hypothesis of local isotropy \cite{kolmogorov1941local} is invoked to simplify the dissipation tensor as follows:
\begin{equation}
    \label{EPS-ISO}
    \varepsilon_{ij} \approx  \frac{1}{3} \varepsilon \delta_{ij} ,
\end{equation}
where $\delta_{ij}$ is the Kronecker delta function and $\varepsilon$ is the magnitude of kinetic energy dissipation. Then,
Kolmogorov's second similarity hypothesis is called upon to develop a transport equation for $\varepsilon$. It is assumed that the small-scale dissipation rate can be approximated as the spectral cascade rate of energy transfer from large to small scales. Thus, the modeled transport equation for $\varepsilon$ can be developed exclusively employing largescale features of turbulence.
The resulting form of the $\varepsilon$ equation is \cite{launder_1972}: 
\begin{equation}
    \label{EPS}
    \frac{ \partial \varepsilon}{\partial t} + \overline{U_k} \frac{\partial \varepsilon}{\partial x_k} = 
       C_{\epsilon 1} \frac{P \varepsilon}{K} -  C_{\epsilon 2} \frac{\varepsilon^2}{K} + 
       \frac{\partial}{\partial x_j} (( \mu + \frac{\mu_t}{\sigma_{\epsilon}}) \frac{\partial \varepsilon}{\partial x_j} ).
\end{equation}
In the above, $P$ is the production of turbulent kinetic energy; $C_{\epsilon 1}, C_{\epsilon 2}$ are dissipation transport closure coefficients, $\mu_t$ is eddy viscosity and $\sigma_{\epsilon}$ is the dissipation Prandtl number.
As mentioned earlier, the unknown coefficients are called the dissipation transport closure coefficients (TCC) and are calibrated using data from  canonical flows:
\begin{enumerate}
    \item First, the coefficient $C_{\epsilon 2}$ is determined from decaying isotropic turbulence data. The asymptotic decay rate of kinetic energy is directly related to this coefficient.
    \item Once $C_{\epsilon 2}$ is known, the  the coefficient $C_{\epsilon 1}$ is determined from fixed-point analysis to yield the correct production-to-dissipation ratio in homogeneous shear turbulence. The right combination of $C_{\epsilon 1}$ and $C_{\epsilon 2}$ is needed to obtain the established value of the production-to-dissipation ratio.
    \item The dissipation Prandtl number $\sigma_{\epsilon}$ is then determined as a function of $C_{\epsilon 1}$, $C_{\epsilon 2}$ and coefficients in the constitutive closure relation to produce the correct log-law behavior in equilibrium flat plate boundary layers.
\end{enumerate}
Thus,  the TCC values and the constitutive closure coefficient (CCC) values are intricately inter-related. Yet, this phenomenological model cannot account for non-equilibrium effects as, in those cases, the spectral cascade rate and dissipation rate can be significantly different. Thus, the mathematical form of the dissipation equation is highly restrictive and precludes non-local and non-equilibrium physics.

\paragraph{Pressure-strain correlation modeling.} The closure model for the pressure-strain correlation term represents another significant challenge as it attempts to represent non-local NMM and elliptic pressure effects with local Markovian closures \cite{mishra_girimaji_2013},\cite{mishra_girimaji_2014},\cite{mishra_girimaji_2017},\cite{PhysRevE.92.053001},\cite{mishra2010} and \cite{mishra2016}. Nearly all popular models  \cite{1975JFM....68..537L},\cite{article} make the following simplifications: (i) the non-local wave-vector effects are represented purely in terms of the local velocity field; and, (ii) the inhomogeneity effects are completely neglected. These assumptions are clearly invalid in flows with coherent structures and rapidly distorted or strongly inhomogeneous flows. Subject to these key simplifications, the pressure-strain correlation in standard models  is expressed in terms of kinetic energy, dissipation, mean velocity-gradient tensor, and Reynolds stress tensor (equation 76 in \cite{speziale1991analytical}):
\begin{equation}
    \label{prstr}
    \Pi_{ij}  = \Pi_{ij} (K, \varepsilon, {\bf b}, {\bf S}, {\bf W}).
\end{equation}
In the above equation, ${\bf S}$ and ${\bf W}$ are the mean-flow strain and rotation rates. Various pressure-strain correlation closure models use different tensor-basis sets. The pressure correlation coefficients (PCC) in the closure expressions may vary among models, depending upon the specific calibration cases. Overall, both the mathematical structure of the model and the calibration rationale are inadequate.

\paragraph{Dynamical System Integration of closures.} 
As mentioned in the previous section, it is vital to ensure that the closure models for individual processes described above are mutually compatible and yield appropriate holistic behavior.
To accomplish this, traditional closure approaches undertake important additional measures:
\begin{enumerate}
    \item Impose mathematical constraints pertaining to classical realizability \cite{pope_2000} and, when possible, comprehensive realizability \cite{mishra_girimaji_2014}.
    \item Ensure consistency with established physical behavior at different extremes of turbulence behavior - e.g., rapid distortion limit \cite{crow_1968}, \cite{girimaji2003}; and, log-layer in wall-bounded flows \cite{wilcox1998turbulence}. 
    \item Perform fixed point analysis to establish that the dynamical system of equations has the correct equilibrium and bifurcation characteristics \cite{article}, \cite{girimaji_2000}.
\end{enumerate}
However, as discussed in \cite{mishra_girimaji_2017}, due to the limitations of the locality assumption,  it is not possible to develop a closure model expression that satisfies all of the above conditions even in the simplest flows. In other words,   Markovian closure models that individually and collectively uphold all of the required characteristics may not exist. The latest developments in the SMC approach are summarized in Hanjalic and Launder \cite{doi:10.1063/5.0065211}.

In summary, the key assumptions invoked for the SMC modeling of the individual terms in the Reynolds Stress Evolution Equation are: (i) mild inhomogeneity; (ii) near-equilibrium turbulence; and, (iii)  no significant spatio-temporal memory (NMM) effects. These shortcomings are overcome to some extent by ensuring that the overall system of equations satisfies key compatibility and consistency conditions leading to a reasonable fixed-point or equilibrium behavior.  However, the inherent deficiencies of the simplifying assumptions continue to adversely affect the model performance in strongly inhomogeneous and non-equilibrium flows, especially in the presence of coherent structures. 

\subsubsection{Two-equation models.}

Two-equation turbulence models can be developed from two different perspectives, both of which lead to a
Reynolds stress constitutive relation dependent upon the mean velocity-gradient field. The first approach is empirical and builds on the analogy to kinetic gas theory. Along the lines of  Newtonian and Fourier transport, the Boussinesq approximation \cite{boussinesq1877essai} postulates
\begin{equation}
    \label{bouss}
    R_{ij} = -2 \nu_t S_{ij} + \frac{2}{3} K \delta_{ij}.
\end{equation}
Here, $\nu_t$ is the eddy viscosity which is specified in terms of turbulence length and time scale. For complex turbulence flows, extended thermodynamic principles can be invoked to include higher-order tensors involving strain and rotation rates:
\begin{equation}
    \label{extbouss}
    R_{ij} = R_{ij} (K, {\varepsilon}, {\bf S}, {\bf W}).
\end{equation}
The coefficients of such a relationship can be found by calibration over canonical flows \cite{huang1996generalized}.

The second and more rigorous approach proposes that the Reynolds stress constitutive relation should be a solution of the transport equation (\ref{RSEE}). For a general flow, the Reynolds stress at a given point in space and time can be represented symbolically as \cite{LUMLEY1979123}
\begin{equation}
\label{nonlocal}
\overline{u_i u_j} ({\bf x}, t) \equiv {\cal F}_{ij} (\nabla \bf{\overline{U}} ({\bf X}, T)), 
\end{equation}
where ${\cal F}_{ij}$ is a functional. The implication is that the Reynolds stress is non-Markovian and depends on the mean velocity-gradient field of the entire space and time domain ($\bf X$, $T$) \cite{LUMLEY1979123}. Such a relation would include all NMM effects but is very difficult to determine. To obtain a manageable constitutive relation, the  Markovian simplification  is invoked:
\begin{equation}
   \frac{D b_{ij}}{Dt} \sim 0; \; \; \mbox{implying} \;\; \overline{u_i u_j} ({\bf x}, t) \equiv F_{ij} (\nabla \bf{\overline{U}} ({\bf x}, t)). 
\end{equation}
Here $F_{ij}$ is a function only of the local velocity gradient field.  
%Rather than model Reynolds stress directly, most approaches seek a closure expression for the anisotropy tensor $b_{ij}$ given in equation (\ref{bij}). 
It is important to note that this is a very restricting simplification that is valid only when the NMM effects are very weak.
Then, {\em and only then}, the general tensor expression for the anisotropy tensor $b_{ij}$ can be expressed using representation theory as \cite{pope_1975}
\begin{equation}
    \label{reptheory}
     b_{ij} ({\bf x}, t) =   \Sigma_{\lambda = 1}^{10} G_{\lambda} T_{ij} ^{\lambda}({\bf x}, t).
\end{equation}
Here, $T_{ij}$ are basis tensors constructed with local strain-rate ($S_{ij}$) and rotation-rate ($W_{ij}$) tenors. The $\lambda$'s are the constitutive closure coefficients (CCC) which are functions of the scalar invariants of the velocity gradient tensor. These coefficients are then determined using the  Algebraic Reynolds Stress Modeling (ARSM) strategy (\cite{Rodi}, \cite{gatski_speziale_1993},\cite{girimaji1996}) -- by requiring equation (\ref{reptheory})  to be a solution of the RSEE (equation \ref{RSEE}) in the weak-equilibrium limit. In assigning the values to the CCC, it is important to preserve the relations with the dissipation transport closure coefficients (TCC) mentioned earlier in the section.
Overall, two-equation models are significantly more restricted in their formal validity than SMC approaches but offer the advantage of lesser computational effort.

%Reynolds stress is first expressed as
%\begin{equation}
%\label{ARSM}
%  R_{ij} = 2K \Sigma_{\lambda = 1}^{10} G_{\lambda} T_{ij} ^{\lambda}({\bf x}, t) + \frac{2}{3} K %\delta_{ij}.
%\end{equation}
%This form of $R_{ij}$ is substituted in the 

In addition to modeling the constitutive relation, the two-equation RANS approach requires solving modeled transport equations for turbulent kinetic energy, $K$, and dissipation, $\varepsilon$ (equation \ref{EPS}). The kinetic energy equation is of the form:
\begin{equation}
    \label{TKE}
    \frac{ \partial K }{\partial t} + \overline{U_k} \frac{\partial K}{\partial x_k} = 
       P  -  \varepsilon + 
       \frac{\partial}{\partial x_j} ((\mu + \frac{\mu_t}{\sigma_{k}}) \frac{\partial K}{\partial x_j} ),
\end{equation}
where $\sigma_k$ is the kinetic energy Prandtl number.
 
\paragraph{Unsteady RANS.} Unsteady RANS (URANS) computations are sometimes utilized to provide some level of scale resolution. While URANS may provide marginal improvement over steady RANS in some cases, it is important to clearly understand the role and limitations of URANS. From a formal perspective, URANS models are meant for statistically unsteady flows, and not for resolving scales in a statistically steady flow. In statistically steady flows,  unsteadiness observed in time-accurate RANS computations may be spurious, arising from modeling inadequacies rather than the underlying flow physics.

%Despite the clear limitations of locality, omission of local higher-order derivatives  and limited tensor basis, the model implied by equation
%(\ref{reptheory}) is routinely used in complex flow calculations wherein the underlying assumptions are violated.
%Many machine learning methods also seek to determine the $\lambda$s in equation (\ref{reptheory}) in complex flows,
%without heed to the inherent limitations.

\paragraph{Takeaway.} Modeling at this level requires a holistic treatment of the full dynamical system aside from individual term closures. Even for the individual closures, 
most models treat the turbulence field as weakly homogeneous. 
Thus, in these approaches, the modeled turbulence field is essentially a piece-wise homogeneous field patched together with elementary gradient transport effects to account for any spatial variations. 
Such treatment cannot adequately account for strong inhomogeneity effects that lead to large-scale instabilities and coherent structures. 
Yet, these models are used to compute complex industrial flows which are strongly non-Markovian due to the presence of coherent structures.
Two-point closure is the lowest level in which some elementary (and by no means all) NMM effects can be formally described.

\subsection{Two-point closures}

In complex flows, the Reynolds stress at a given point in time and space depends upon the temporal history and spatial distribution of the flow field (Lumley \cite{LUMLEY1979123}). 
Two-point closures describe the evolution of Reynolds-averaged one-time, two-point correlations which can potentially capture some elementary non-local effects (\cite{bertoglio:hal-00241150}, \cite{arun_sameen_srinivasan_girimaji_2021}). The central statistic under consideration is the one-time two-point velocity correlation given by
\begin{equation}
    \label{phiij}
    \phi_{ij} ({\bf x}, {\bf r}, t) \equiv \overline{u_i({\bf x}, t) u_j({\bf x + r}, t)}
\end{equation}
In the above, ${\bf x}$ is the location of interest and ${\bf r}$ is the separation vector of all other points in the flow domain. 
It is important to note that in inhomogeneous flows, any two-point statistic is a function of not only the location of interest but also of the separation distance in three-dimensional space. Thus, ${\bf x}$ and ${\bf r}$ belong to the entire flow domain and all two-point correlations reside in a six-dimensional space. Consequently, the computational burden of a two-point closure method is significantly higher than that of a single-point RANS model.

The evolution equation for $\phi_{ij}$ consists of functions of several higher-order two-point correlations
 (\cite{arun_sameen_srinivasan_girimaji_2021}, \cite{PhysRevFluids.3.124608}). Some of the physical phenomena governing the two-point correlation evolution are conceptually similar to those in single-point Reynolds stress evolution: production, turbulent diffusion, turbulent transport, and pressure effects. The major novelty is that these phenomena now involve two-point physics, thereby explicitly accounting for some non-local effects. For example, the production term is of the form 
\begin{eqnarray}
    \label{prod}
    P_{ij}^{\phi} ({\bf x}, {\bf r}; t) & \sim & - \phi_{ik} ({\bf x}, t) \frac{\partial 
    \overline{U_k}({\bf x}, t)}{\partial x_j} -  \phi_{ik} ({\bf x}, t) \frac{\partial 
    \overline{U_k}({\bf x + r}, t)}{\partial x_j} \\ \nonumber
   && 
   - \phi_{kj} ({\bf x}, t) \frac{\partial 
   \overline{U_k}({\bf x}, t)}{\partial x_i} -  \phi_{kj} ({\bf x}, t) \frac{\partial 
   \overline{U_k}({\bf x + r}, t)}{\partial x_i}.
\end{eqnarray}
This expression accounts for the production at a location ${\bf x}$ due to the mean velocity gradient at (${\bf x+r}$). Along the same lines, the pressure correlation terms account for the influence of the entire flow domain on the point of interest.  Similarly, the two-point transport term describes the transport at a given location due to the influence of fluctuations at another location. The pressure effects, turbulent transport and dissipation of two-point correlations require closure modeling.

Transport equations of two-point correlations also contain novel phenomena not encountered in one-point closures: extra source/sink due to inhomogeneity, interscale transfer due to spatial variations in the mean velocity field and turbulence statistics \cite{arun_sameen_srinivasan_girimaji_2021}. The inhomogeneous mean velocity term incorporates the effects of hydrodynamic instability directly into the evolution equation. The interscale transfer terms represent the cascade effect and require closure modeling. While the physics-based benefits of these two-point phenomena are evident,  key challenges remain:
\begin{enumerate}
    \item The computational burden is significantly higher than single-point RANS, especially in complex flows.
    \item Closure modeling of two-point phenomena pertaining to turbulent transport, pressure effects and other phenomena are of a much higher degree of complexity than their single-point counterparts -- e.g., two-point NMM effects.
    \item Many of the two-time correlations important in unsteady flows are still unaccounted. 
\end{enumerate}
In order to reduce the computational burden of two-point closures, some authors have developed one-point models derived from two-point closures by making simplifying assumptions (\cite{kassinos_reynolds_rogers_2001}, \cite{PhysRevFluids.3.124608}). While the theoretical premise is promising, the practical utility of these methods is yet to be clearly established in the literature.

\paragraph{Takeaway.} Current two-point closure approaches possess some intellectual merit but have shown limitations in practical applications. Important unsteady flow features still fall outside the scope of the model. Many researchers in the field hold the perspective that the challenges in closure modeling and the computational effort required for two-point models may not be commensurate with the overall benefits of the approach. An argument can be made that, for a comparable level of computational effort, simulations using filtered Navier-Stokes equations may offer greater benefits.

\subsection{Filtered Navier-Stokes equations}

The physical concepts and mathematical framework for closure modeling of subgrid stress ($\tau_{ij}$) are similar to that of the Reynolds stress $R_{ij}$. However, the degree of difficulty of closure modeling will be significantly simpler if all of the coherent structures do indeed reside in the resolved scales $\tilde{U}_i$. Under these conditions, the unresolved scales can be taken to be Markovian. As seen previously, turbulence features that do not possess significant memory effects lend themselves to fairly straightforward closure modeling.

The filtered Navier-Stokes methods can be classified into two broad categories - large eddy simulations (LES) and scale resolving simulations (SRS). In general, the classical LES and its variants resolve a significant portion of the spectrum even in flow scales wherein only stochastic turbulence is present
(\cite{smagorinsky1963general}, \cite{DynamicSGS}, \cite{lesieur1996new}, \cite{Pope_2004}). The LES subgrid stress is typically specified as an algebraic function of the resolved flow field and may not require the solution of any other transport equation.  In principle, SRS methods seek to resolve a minimal range of coherent scales and model the rest, thereby achieving a more favorable balance between accuracy and computational cost.
The SRS methodology consists of two approaches - hybrid and bridging. A hybrid turbulence approach combines LES and RANS within a flow simulation (\cite{shur2008hybrid}, \cite{heinz2020review}).  Regions of flow characterized by coherent structures are computed with LES. A standard RANS model is used in other regions wherein the turbulence field does not exhibit any non-local features. A bridging SRS model uses the same turbulence closure in all regions of the flow, but the model coefficients are varied to change the resolution in a manner consistent with closure physics (\cite{girimaji2006partially}, \cite{girimaji2006partiallyb}, \cite{schiestel2005towards}, \cite{chaouat2005new}).  As more scales of motion are modeled, the closure modeling in the bridging-SRS context is not as simple as in LES but is significantly easier than in the RANS approach. 

\paragraph{Takeaway.} Filtered Navier-Stokes methods enjoy wide usage in the research community, but are still sparingly used for large-scale industrial computations due to the large computational requirements. The SRS methods need further development before they can be considered front-line design tools.

\section{Machine learning for turbulence modeling}

Thus far in the paper we have identified turbulence features that pose closure modeling challenges (Section 3) and summarized the deficiencies of different traditional turbulence modeling approaches (Section 4). 
Given the success of machine learning in other fields, it would appear that ML techniques may possess the ability to transform the field of turbulence closure modeling.
In this section, we will examine the potential of ML methods to overcome the unique challenges of turbulence closure modeling. 
Toward this end, this article seeks to answer the following questions: 
\begin{enumerate}
\item When and why do traditional RANS closures fail?
\item How is turbulence closure modeling different from other problems where ML has enjoyed success?
\item What are the physics-related challenges to developing generalizable  ML-based closure models?
\item What are the data requirements for developing generalizable ML-based models? Is it feasible to generate the required data?
\item What are key lessons learned and meaningful future research directions? 
\end{enumerate}
Despite its considerable limitations, the one-point RANS approach continues to be widely used for industrial design and development applications due to its computational simplicity.  Therefore, much of the following discussion is centered around the RANS approach.

\subsection{When and why do traditional RANS closures fail?}

As a prelude to developing ML-based improvements, it is critically important to first understand when traditional turbulence models work and why they fail under other circumstances. 
The success of traditional turbulence models is assessed in terms of their ability to (i) reproduce critical features of important benchmark flows; and (ii) adequately predict unseen flows without too much {\em ad hoc} modification.
Following the discussions in Sections 3 and 4, it is beneficial to examine the amenability of the different types of turbulence fields to closure modeling. 

\paragraph{Equilibrium Stochastic Homogeneous Turbulence.}  The one-point closure paradigm and corresponding simplifying assumptions are ideally suited for near-equilibrium homogeneous flows. As the dynamical system is close to an equilibrium state, the interplay between the different equations can be established using a fixed-point analysis \cite{girimaji_2000}. 
The overall effect of the fluctuating field on the mean flow is that of enhanced viscous-diffusive transport.  The turbulent kinetic energy cascades unidirectionally from large to small scales - the spectral cascade rate is thus equal to dissipation. From an energy dynamics viewpoint, turbulence merely serves to dissipate kinetic energy. Consequently, the flow physics also allows for a local (Markovian) constitutive relation.  When properly calibrated, traditional models can represent the underlying physics accurately. In fact for these flows, RANS performs quite well.
%{\em In principle, kinetic theory arguments for an equilibrium thermodynamic system can be extended to these turbulent fluctuations. More formally, a model for Reynolds stress can be obtained as the fixed point of the SMC equations. However, constitutive relationship between the mean velocity gradient field and turbulence statistics can be complicated due to pressure effects (\cite{mishra_girimaji_2013},\cite{mishra_girimaji_2014},\cite{mishra_girimaji_2017},\cite{PhysRevE.92.053001},\cite{mishra2010},\cite{mishra2016}CITE Mishra and Girimaji papers. Indeed, there can strong bifurcation in the behavior as a function of the mean velocity-gradient parameters (\cite{PhysRevE.92.053001}Mishra \& Girimaji, Hydrodynamic stability of three-dimensional homogeneous flow topologies Phys. Rev. E 92, 053001, 2015). Stochastic fluctuations in flows with added body forces can also can also exhibit bifurcation as a function of the extra strain-rate parameter. Overall, while constitutive models for this approach are possible, the relationships can difficult to derive or develop.}

\paragraph{Non-equilibrium Stochastic Homogeneous Turbulence.}
The traditional methods are less successful in these flows. The history (memory) effects are important, and there can be a time lag between the spectral cascade rate and dissipation, invalidating the fundamental assumptions underlying the dissipation equation. Furthermore, the local constitutive relation can be elastic rather than viscous, as the stress depends on the total deformation rather than the current deformation rate. The use of a Markovian viscous relationship in elastic constitutive regimes can lead to the so-called realizability violations. Current methods use limiters to overcome realizability violations, but the underlying physics can be significantly compromised. In this regime, traditional models may be adequate, but only if constructed by suitably accounting for the memory/history effects.

\paragraph{Flows with strong coherent structures.}
In the presence of large coherent structures, the advective stirring action of the fluctuating field is more dominant than viscous or diffusive transport effects. Importantly, the reversibility of advective action can lead to an inverse energy cascade. Consequently, dissipation can be completely different from the spectral cascade rate, invalidating many of the simplifications of canonical turbulence. Clearly, from a modeling perspective, the spatio-temporal non-Markovian effects due to coherent structures are critical. In general, traditional Markovian RANS models perform very poorly. While tuning the coefficients can improve predictions of some statistics, the overall model performance can be very poor and, perhaps, catastrophically wrong for unseen flows.
 
\paragraph{Takeaway:} By their very construct, one-point closures (RANS and SMC) are best suited for the viscous-diffusive action of the fluctuating field. They are generally effective for near-homogeneous flows dominated by stochastic turbulence. In more complex flows, the non-universality of large-scale turbulence, especially, the presence of coherent structures, renders these models ineffective. The main reason for the failure of these models is the inappropriate use of near-equilibrium Markovian models in flows with significant non-Markovian memory (NMM) effects. From a flow physics point of view, the failure can be attributed to the use of irreversible diffusion models for capturing reversible advective effects.

\subsection{How is turbulence different from other problems where ML has been successful?}

Machine learning algorithms have been remarkably successful in areas such as speech recognition, computer vision, automation of ordered systems, and some areas of physics. In these problems, the main role of ML is to find a reasonable mapping function between input and output.  Even when a local constitutive relation is feasible, turbulence closure modeling presents a different set of challenges than the areas in which ML methods have found success.

As seen in Section 2, turbulence modeling - especially at the RANS level - entails modeling several unclosed terms in a  dynamical system of multiple partial differential equations.
{\em The successful closure of individual terms using input-out mapping function is a necessary, but not sufficient condition for an accurate model computation.} The individual input-output mapping must also lead to acceptable behavior of the dynamical system in a holistic sense.
%The RANS equations constitute a complicated dynamical system comprising of multiple closure models.
%A  predictive RANS computation entails simultaneous solution of a system of multiple interacting partial differential equations and algebraic equations.   
In traditional RANS approaches, the interplay among the models is carefully orchestrated using a fixed-point analysis of the entire dynamical system. Thus, the transport closure coefficients (TCC) and constitutive closure coefficients (CCC) are strongly coupled to preserve key physical and mathematical properties of the overall system \cite{Taghizadeh_2020}. At the current time, there are no appropriate ML tools (known to the author)  for holistically and collectively training all model coefficients simultaneously to yield the desired fixed point behavior. Instead, currently
ML is used only to train the coefficients (CCC) of the algebraic constitutive closure using high-fidelity data and TCC values are left unchanged. Modifications to CCC without corresponding changes to TCC can result in physical inconsistency and mathematical incompatibility issues, leading to gross inaccuracies and even unphysical results \cite{Taghizadeh_2020}.

\paragraph{Mathematical Compatibility:} Any unilateral changes to CCC can affect the fixed-point behavior of the RANS dynamical system. Attracting fixed points of the traditional model system can potentially change to limit cycles or even repellers.  Such a transformation of the dynamical system can significantly affect the stability and convergence characteristics of the calculations. Even in trained flows, a stand-alone RANS calculation with new CCC coefficients may not reproduce the same mean flow because of the incompatibility between CCC and TCC. 

\paragraph{Physical Inconsistency:} Apart from changing the mathematical character of the model dynamical system, any unconstrained changes to CCC based purely on some training data can also affect the physical underpinning of the closure model. For example, the traditional model CCC and TCC are jointly constrained to yield the correct log-law behavior in a boundary layer. This may be easily compromised if one set of coefficients are modified using ML and the other set is not.  Further, the CCC changes can affect the fixed-point anisotropies in previously calibrated flows. The modeled dynamical system of equations must possess and preserve several key physical and mathematical features including  Reynolds stress realizability, dynamic realizability \cite{girimaji_2004}; and consistency with different asymptotic limits such as rapid distortion theory \cite{mishra2010}, \cite{girimaji2003}. These constraints arise from the requirement that the unresolved flow field satisfy the conservation of mass and momentum equations. Failure to satisfy these constraints would be tantamount to violating key physical laws. 

\paragraph{Takeaway:} Traditional approaches model the Reynolds stress constitutive relation as an integral part of a larger dynamical system to ensure overall reasonable prediction. Most ML approaches only seek `isolated' best-fit constitutive coefficients without any regard for other dynamically coupled equations. Not accounting for the dynamic interactions between different elements of the closure model can lead to poor {\em a posteriori} performance even in trained flows.
%Many of the current ML-RANS approaches involve unconstrained and unilateral changes to CCC based on training data. Such practice can do harm to the generalizability of the model, while possibly improving performance in a small subset of trained flows.
%There is a clear need for careful assessment of generalizability to unseen or untrained flows. 
A  reasonable strategy for dynamically coupling CCC and TCC values must be established. Taghizadeh et. al
\cite{Taghizadeh_2020} propose a closed-loop training strategy that addresses the coupling between CCC and TCC.  Thus, even when physics permits local constitutive relations, further developments are needed to effectively model the overall dynamical system holistically, rather than consider the constitutive equation in isolation.

%\begin{enumerate}
%    \item  The ML model is a part of a dynamical system which consists of multiple transport equations with other closures. Thus the ML model not only must produce reasonable constitutive relation, but the overall system behavior must yield appropriate fixed attractors. Thus the compatibility between ML elements and other closures must be appropriately orchestrated. 
%    \item Turbulence closure models must possess and preserve several key physical and mathematical features including  Reynolds stress realizability (REFERENCES), dynamic realizability (Girimaji, JFM, 2004); and consistency with different asymptotic limits such as rapid distortion theory (Mishra and Girimaji, Flow Turbulence and Combustion, 2010; Girimaji, Jeong and Poroseva, POF, 2003). These constraints arise from the requirement that the unresolved flow field satisfy conservation of mass and momentum equations. Failure to satisfy these constraints would tantamount violating key physical laws.
%\end{enumerate}
%In recent works of Taghizadeh et. al (2022), it is shown that some of the above challenges can be overcome with a closed-loop training approach. This approach will require more computational effort than the standard open-loop training method, but will ensure reasonable compatibility between different elements of the closure models and be more consistent with the underlying flow physics. 

\subsection{Challenges to developing individual ML closures.}

Aside from the holistic treatment, there are other challenges encountered in ML modeling of individual unclosed terms.  While neural networks can potentially approximate any input-output mapping function, the size of the network can be infeasibly large and generalization may not be straightforward \cite{Goodfellow-et-al-2016}. Further, it has been shown that finding a suitable network architecture for achieving prescribed accuracy for a given application can be an intractable problem \cite{https://doi.org/10.48550/arxiv.2205.13531}, \cite{kutynoik}. 
Based on these observations, we propose the following as the critical characteristics that determine the success of neural networks in high-dimensional (large number of degrees of freedom) problems such as turbulence:
\begin{enumerate}
\item Generalizability of underlying physics, enabling models developed in some parameter regimes to be applied to data in other unseen regimes.
\item Existence of a low-dimensional manifold in the input-output function space, ensuring a manageable network size.
\item Ability to identify the optimal network architecture when a low-dimensional manifold is indeed present.
\end{enumerate}
The above criteria must be viewed as necessary but not sufficient conditions for the success of neural networks in the field of turbulence modeling. 

 %   \item The solution (input-output mapping function) must occupy a low-dimensional manifold in a higher-dimensional space. It the mapping has a large number of degrees of freedom, the size of the network and the data required for training can be infeasibly large.  
  %  \item The parameters characterizing the low-dimensional manifolds must be tractable or cannot be used as features. It the parameters are not observable or computable for them to be of practical utility. 
 %   \item Sufficient data must be available throughout the parameter space of turbulence physics.
  %  \item Clear guidance for selecting the network architecture must available.
%\end{enumerate}

\paragraph{Generalizability of turbulence physics} 
%The statistical tools constituting the neural networks are largely developed on the presumption of independent identically distributed (iid) variables. 
{\em It is reasonable to presume that generalizability is achievable only if the physics underlying the input-output mapping function is self-similar (universal) and the appropriate similarity variables that characterize the function are utilized.}  
%As has been pointed out earlier, turbulence large scales can be highly flow dependent and hence not universal. The largescale self-similarity can vary significantly from flow to flow, especially in the presence of coherent structures. 
Kolmogorov theory accurately describes the self-similarity of inertial and smallscale turbulence features and identifies the relevant similarity parameters. Intermediate and smallscale features are likely amenable to a generalizable ML model.

The central object of interest in this study, the Reynolds stress, is determined mostly by the turbulence largescales. Even in the simplest of non-trivial turbulent flows, the largescale features are far from universal or self-similar. %Indeed, the largescale flow physics and the parameters that describe the flow can be quite different.
In the simplest statistically two-dimensional homogeneous flows, the type of instability in rotation and strain-dominated flows are very different \cite{PhysRevE.92.053001}. In strain-dominated flows, inertial effects are destabilizing while pressure is stabilizing. Rotation-dominated flows exhibit parametric resonance instability wherein pressure action destabilizes an otherwise stable flow. Further, diverging and converging three-dimensional flows exhibit very different statistical characteristics. The equilibrium anisotropy level is dictated by inertial or pressure effects depending upon several factors. 
In a recent paper \cite{Tag2021}, it is shown that ML models trained in one equilibrium homogeneous turbulence flow domain cannot accurately predict the equilibrium states in other regions of the homogeneous turbulence parameter space. In non-equilibrium turbulence, inhomogeneous flows and flows with coherent structures, generalizability can be even more challenging. Further, the non-locality effects are highly geometry-dependent. In many flows, the coherent structures can change significantly with increasing Reynolds numbers \cite{doi:10.1146/annurev.fl.28.010196.002401}.  Additional influences such as body forces, system rotation, streamline curvature, and compressibility can add more bifurcations to an already complicated system.
Therefore, we suggest that the generalizability of largescale turbulence features (such as Reynolds stress) is likely to be very challenging.

\paragraph{Existence of low-dimensional manifold.}
In spatio-temporal phenomena such as turbulence, the analysis of low-dimensional manifolds (or reduced data manifolds) is inherently difficult \cite{https://doi.org/10.48550/arxiv.2108.09545}. For the case of statistically homogeneous turbulence, wherein the statistics are functions of time only, \cite{2001ThCFD..14..259G} examines the solution manifold around the equilibrium state. It is shown that an attracting low-dimensional manifold does exist and it can be characterized reasonably accurately. However, in the rapid distortion state of homogeneous turbulence, a low-dimensional representation is not possible due to the strong dependence on the initial condition. Further, in flows with coherent structures, dimensionality reduction can be an intractable problem due to complex spatiotemporal phase-space behavior.

\paragraph{Network architecture optimization.}
The performance and generalizability characteristics of a neural network is profoundly dependent on the architecture and the hyperparameters. 
%Thus it is vital importance to seek out optimal choices for different applications.
In fields such as image processing, automated strategies that optimize neural network architecture without user intervention are being pursued \cite{luo2018neural}. However, turbulence closure modeling poses a different set of challenges and there is no clear consensus on what constitutes an optimal network architecture. Networks with a wide range of layers and nodes have been reported in the literature.
Research on developing a formal procedure for characterizing and identifying optimal neural network architectures for turbulence modeling applications would be of much use.

\subsubsection{Non-local models} 

%Evidently local one-point models are not easily generalizable due to the non-universality of largescale structures. 
Some authors suggest that the generalizability limitation can be overcome with ML models that use non-local information - i.e., ML closures that incorporate data from a region around the point of interest. However, these non-local models encounter their own set of challenges. 

In the one-point RANS (local effects) model context, the ML-based methods are based on the reasonably rigorous foundational framework of the traditional physics-based models. For incorporating non-local effects in the ML methods, it would be beneficial to build on the foundational framework of the two-point closure approach. The development of data-driven neural networks for two-point statistics is a problem of significant conceptual and practical difficulty. There have been some efforts to develop nonlocal data-driven models in spectral space for simple flows  \cite{doi:10.1063/5.0064394}. These models perform well for simple flows, but the spectral treatment is not possible for most flows of general engineering interest.

Some authors have attempted a markedly different data-driven nonlocal approach \cite{ZHOU2021113927}. 
These authors interpret equation \ref{nonlocal} in its most elementary form and postulate:
\begin{equation}
    \label{nonlocalbij}
      b_{ij} ({\bf x}) = {\cal F} ({\bf\overline U} ({\bf X}))
\end{equation}
where ${\bf U}$ is the mean velocity field, ${\cal F}$ is a functional and ${\bf X}$ represents the entire flow domain. In principle, a neural network for ${\cal F}$ can be determined using data over the entire flow domain. 
However, in practice, using convolution over the entire flow domain is impractical, and hence the region of nonlocal influence is restricted to a specified area around the point of interest.
%Purely local (single-point) closures is clearly a permissible solution of this approach. 
Potentially, this approach can lead to improvement over local RANS models. 
However, there are many practical challenges:
\begin{enumerate}
    \item Many of the non-local effects may not be completely described in terms of the two-point correlations. Only the NMM effects incumbent in two-point one-time statistics are included in this approach.  Despite the use of spatially nonlocal mean velocity field information, this approach lacks the formal foundation of physics-based two-point approaches which explicitly capture relevant physical phenomena. It is unclear how physical phenomena that manifest through two-point correlations of pressure and fluctuation velocity gradients highlighted in the previous section will be addressed. Indeed, these two-point pressure-strain correlation terms could be of much importance in many complex flows.
    \item In unseen flows, the region of dominant influence (range of ${\bf X}$) is not known {\em a priori}. 
    %Thus, further simplifications must be invoked to reduce the region of influence to manageable levels. 
    Even within a given flow, the region of influence can change from point to point. Thus, for each computational point, the optimal convolution domain is different.
    \item Generalizability of non-local closures will be even more challenging than local closures. This is because two different flows may not have similar non-local structures and hence lessons learned from one flow will not carry over to another.
    \item It will be seen shortly that the data requirement of a sufficient number of independent samples of statistically similar regions (in terms of ${\bf U}$ distribution) can be very difficult to accomplish.
\end{enumerate}

\paragraph{Takeaway:} Largescale turbulence physics is strongly flow-dependent (non-universal) and hence not easily amenable to generalizability. Traditional models are based on mathematical and physical principles which impart some degree of generalizability. 
%While these models may not yield accurate results in complex flows, they generally do not fail catastrophically. 
Current ML turbulence closures do not necessarily impose the first-principle requirements. {\em There is clear evidence that training on data alone, without other physics-based or mathematical intervention, does not lend itself to generalizability} \cite{Tag2021}. Lacking more precise physics-based guidance and constraints, catastrophic failure of ML models in unseen flows cannot be ruled out.

\subsection{Data requirements and feasibility of generating needed data}

A turbulence closure model provides a mapping between known resolved velocity field variables and various unclosed statistics. 
It is evident from the foregoing discussion that ML models may be limited in their ability to accurately compute unseen turbulent flows - in other words, ML models cannot effectively extrapolate. This implies that ML models must be trained in all regions of the parameter space. However, the parameter space of all possible states of the resolved turbulence field is unbounded. 
Some authors advocate training the models over a set of benchmark flows to improve the predictive capability in unseen flows. Without guidance from first principles, such an approach may compromise the accuracy of the model in each of the benchmark flows without guaranteeing improved performance in unseen flows. 
Therefore, it is vital to train the model with data from all possible domains of turbulence physics likely to be encountered in a practical computation. The volume of data required and our ability to generate the requisite high-fidelity data will depend upon the regime of turbulence and type of model (local or non-local).

%Local models:
%-  Stress at a point depends only on the local strain field. Small stencil size, fewer parameters to tune
%ML cannot extrapolate reliably 
%-- data needed from all bifurcation branches.

%Even for homogeneous 2D mean flows, this is a tall order

%Non-Local models 
%-- Stress at a point depends on strain field over a large domain
%Large stencil size, large number of parameters to tune, need significantly more data
 
%-- Large quantities of data from each structure type 
%Many coherent structure types, strongly dependent upon flow geometry
%Unbounded set  of coherent structures 

%-- Unbounded need for training data
%-- For transient coherent structures 
%-- Need time label (dependence) as well
 
\paragraph{Stochastic Equilibrium Turbulence:}   

This most elementary state of homogeneous turbulence generally permits a local (Markovian) constitutive relationship between mean-field strain (velocity gradient field) and Reynolds stress. The mean velocity-gradient parameter space can be characterized in terms of
\cite{das_girimaji_2020}: (i) magnitude of velocity gradient - $\frac{\partial \overline{U_i}}{\partial x_j}\frac{\partial \overline{U_i}}{\partial x_j}$; and (ii) internal structure -  all possible combinations of strain and rotation tensor orientations. For two-dimensional homogeneous turbulence, there is sufficient DNS data for model development \cite{girimaji_2000}. However, for three-dimensional mean flows, very little data is available except for axisymmetric contraction and expansion. In fact, very little is known about all possible homogeneous turbulence regimes. All bifurcations must be identified and a sufficient number of samples in each domain must be gathered.  It is unclear if experiments or DNS can be designed (notwithstanding the expense)  to obtain the required data for arbitrary combinations of strain and rotation rates.  In summary, even in this simplest form of turbulence, the parameter space is large and data is currently available for only a small portion of the domain. Despite millions of hours of computational time, even simple cases such as isotropic turbulence and channel flows are not fully characterized.  Thus generating high-fidelity data for the entire homogeneous-flow parameter space may be an insurmountable challenge.  

\paragraph{Non-equilibrium stochastic turbulence:} 
%The mean-strain state space can be considered to be bounded between decaying turbulence ($Sk/\varepsilon \sim 0$) and 
%the rapid distortion state ($Sk/\varepsilon \gg 1$). 
The parameter space of these flows is even larger since the flow variables depend on initial conditions and transient effects.
Yet again, it is very difficult to set up experiments (or DNS) even for moderately difficult flow configurations. Thus, generating reliable data to train models can be prohibitively expensive. 

\paragraph{ Coherent structures.} These flows are characterized by strong non-local effects. Different locations within a coherent structure can experience vastly different flow physics.
For the same local strain rate, the stress can be vastly different due to underlying flow instabilities. The domain of influence is a strong function of flow-type and can change drastically with flow geometry and Reynolds number. Each sample data unit comprises a three-dimensional flow field, rendering the data requirement inconceivably large.  
%The system may or may not have an attractor. Existence of a generalizable constitutive relation is not guaranteed. Even if it exists, it may not be unique (multi-valued function). Even if it is unique, it may be unknowable. It is unclear how much data is needed to train for a function that is multi-valued and unknowable. is an ML model reliable under these circumstances? Each coherent structure has a different domain of influence
Training a local model over different non-local effects will compromise the validity and utility of the model in simple flows.
%Need to introduce extra features to distinguish between different flows
%But, extra features will add significantly to training efforts.

%\begin{enumerate}
   % \item Inability to extrapolate. Need infusion of physics external to data. We have not been able to do this in traditional modeling for decades. 
    %\item Large parameter space with multiple bifurcations even for homogeneous turbulence
    %\item Unknown bifurcations and multi-point dependence for coherent structures
%\end{enumerate}
%Thus, the ML model has to trained in all possible scenarios that a practical flow is likely to encounter!

\paragraph{Takeaway.} The turbulence-state parameter space is unbounded and even the qualitative behavior of turbulence throughout this space is unknown. 
It is now evident that ML models cannot extrapolate effectively. Gathering high-fidelity data that covers all possible states of turbulence, in the presence of NMM effects, is likely impossible to achieve even in the simplest incompressible flow case.

\subsection{Lessons learned and potential directions for meaningful progress}

It is too early to speculate on the degree of impact that ML will have on turbulence closure modeling. While ML is a powerful tool, it is yet to find its appropriate role in turbulence predictions. Based on the current status of the field, the following inferences can be drawn:
\begin{enumerate}
    \item One of the main reasons for the failure of turbulence models is the non-universal nature of large scales,  especially the non-Markovian character arising from non-equilibrium effects and coherent structures, that cannot be accounted for accurately in the current closure framework.
    \item The sheer volume of data required for developing generalizable (predictive) ML turbulence models in the presence of NMM and non-equilibrium effects is very large.  The generation of requisite data for all possible states of turbulence that may be encountered in industrial flows is unfeasible. 
    \item Due to the non-local NMM effects and data requirements,  a `one-size-fits-all' ML-RANS turbulence model seems unlikely.
    \item Many  ML modeling approaches in current practice disregard important physical principles and mathematical constraints further limiting any possibility of generalizability. Indeed, these practices can set the back progress made by traditional approaches over the last few decades.
\end{enumerate}

Despite the fact that the current rate of progress is much slower than initially projected by some, it is vital to continue exploring different ways of utilizing ML for turbulence closure modeling. 
It is becoming evident that a better synergy between turbulence physics and ML techniques is absolutely vital.  Specifically, any future work must acknowledge the lack of universality of flow physics due to NMM effects and device means to address them in a satisfactory manner.
%It is vital to revisit turbulence physics and characterize it from a ML viewpoint. 
We propose a partial set of topics that may contribute towards this end.
\begin{enumerate}
    \item Formulate a set of quantitative metrics to clearly identify and isolate coherent structures from stochastic (memory-less) turbulence.  
    \item Develop procedures to categorize the types of NMM effects caused by different classes of coherent structures. 
    \item Conceive ways of incorporating information of coherent structures and their NMM effects into physics-informed or inspired neural networks to overcome current limitations.
    \item Formulate ML-friendly physical principles and mathematical constraints that can considerably improve the prospects of generalizability of ML models and reduce dependence on data for supervised training. Consistency with linear theory, rapid distortion theory, and realizability are some considerations that can lead to improved models without the need for excessive data.
    \item Re-imagine SRS as a physics-inspired ML approach that can efficiently simulate complex flows without too much prior data. With the advent of quantum computing and other novel approaches, computing power will continue to grow making selective scale resolution affordable even for complex flows.  The main advantage of SRS is that the accuracy of NMM carrying large scales is ensured as they are resolved. A local (Markovian) ML subgrid model can be used with confidence for the stochastic unresolved scales of motion. 
    Indeed, methods that judiciously combine the strengths of ML and SRS (referred to as ML-SRS)  are likely to be the preferred approaches of the future.
\end{enumerate}

\section{Conclusion} 

The overarching objective of turbulence models is to provide reliable and adequately accurate predictions of complex practical flows at reasonable computational costs. 
For real-world flows, the RANS (Reynolds-averaged Navier Stokes) approach continues to be the method of choice due to its computational simplicity.
However, conventional RANS models fall short in the accuracy department due to the intricacies of the turbulence phenomenon. In recent times, practitioners have sought to enhance turbulence model capability by incorporating data-driven (supervised learning) techniques.   
However, the progress toward a broadly applicable and accurate ML-RANS model has been very slow.
It would appear that the initial expectations for ML to revolutionize the field of turbulence closure modeling may have been too optimistic. 
 Indeed, there are examples wherein the ML-RANS models are inferior to traditional models in unseen flows \cite{rumsey2022nasa}. This article examines the physics underlying closure modeling challenges to assess the inherent limitations of ML-RANS and establish realistic expectations.   Although most of the discussion presented here is restricted to supervised learning and  RANS closures, the considerations also apply broadly to all data-driven turbulence closures.

The three key steps of turbulence closure model development are (i) Filtering (or truncation of the scales of motion) of the governing equations leading to unclosed reduced-order description; (ii) Postulation of the mathematical forms for the unclosed statistics to close the equations describing the reduced-order system; and (iii) Determination of closure model coefficients for broad applicability. To develop a clear comprehension of ML-based turbulence closure capabilities and limitations, we examine the differences between traditional and ML techniques at each stage of model development.

\paragraph{\bf Step 1: Scale truncation.} At high Reynolds numbers, the Navier-Stokes equation exhibits a high degree of freedom in the form of broad spatio-temporal spectra. For the sake of manageable computations, order reduction is carried out by filtering part or the entirety of the fluctuating field. The physical effects of the discarded scales on the retained ones manifest via unclosed statistics representing the eliminated scales. It is crucial to recognize that the very act of order reduction introduces significant limitations. As extensively debated in the field of Statistical Mechanics, the most notable limitation is the loss of reversibility (e.g., Loschmidt paradox and Poincare recurrence \cite{poincare1890probleme}). In reality, the eliminated scales of motion can have a reversible advective effect on the resolved flow field. Yet, the statistical models predominantly have a diffusive character that irreversibly follows the entropy arrow of time.  This irreversibility limitation can have substantial consequences in flows with large coherent structures. Subject to irreversibility limitation, statistical turbulence closure models seek to develop the best possible model.
    
\paragraph{\bf Step 2. Mathematical form of the closure expression.} The next step in the turbulence modeling sequence is the postulation of the mathematical forms for the unclosed statistics to achieve a closed-form description of the reduced-order system. In principle, an unclosed statistic can be expressed as an integral function (a functional) of the resolved flow variables, as shown in equation \ref{nonlocal}. However, a closure model involving a {\em functional} is not of much practical utility due to the complexity of its computation. Practical considerations dictate the use of a more manageable {\em function} that is more local in time and space. Even if some non-local information is incorporated, there is no clear strategy to ensure all the memory effects are included correctly. Generally, non-Markovian memory effects are disregarded, and a near-equilibrium simplification is invoked to postulate a viable closure expression (e.g., equation \ref{reptheory}). Such closure expressions include tunable coefficients or parameters. Thus, the process of arriving at a manageable mathematical form for the closure expression imposes further limitations on the applicability and accuracy of the turbulence model.

With few exceptions, the first two stages of closure model development are identical in traditional and ML-based approaches. The irreversibility limitations introduced by statistical representation and the neglect of non-Markovian memory effects are unlikely to be completely overcome by using ML methods.
At the end of the second stage, the turbulence closure equations constitute a dynamical system with multiple closure coefficients that must be optimally calibrated to yield the best possible behavior across a range of flows. As emphasized in Section 3, the holistic behavior of the system takes precedence over the accuracy of any individual closure model.
In the third stage of model development, closure coefficients are determined using a combination of hypothesis-based considerations and experimental (laboratory and numerical) data. The primary distinction between the two approaches becomes apparent at this stage.

%Barring few exceptions, the first two stages of closure model development are identical in traditional and ML-based approaches. The irrevrsibility limitations introduced by statistical representation and neglection of the non-Morkovian memory effects are unlikely to be completely overcome by using ML methods. 
%At the end of the second stage,  the turbulence closure equations constitute a dynamical system of equations with multiple closure coefficients that must be optimally calibrated in order to yield the best possible behavior over a range of flows.  As pointed out in Sections 3, the holistic behavior of the system of equations is more important than the accuracy of any one closure model.
%In the third stage of model development, the closure coefficients are determined using a combination of hypothesis-based considerations and experimental (laboratory and numerical) data. 
% The main distinction between the two approaches manifests at this stage.
%Complex closure models with multiple coefficients can potentially be more applicable over a broader range of flows than simpler models with fewer coefficients.

\paragraph{\bf Traditional Approach Step 3: Coefficient determination.} The traditional approaches prioritize theoretical considerations, relying on data only as a last resort for calibrating closure model coefficients. To begin with, the behavior of turbulence in various limiting scenarios — such as the rapid distortion limit, equilibrium state of homogeneous turbulence, decaying turbulence, and near-wall asymptotics — is examined to determine or limit some unknown coefficient values. Kinematic constraints, such as realizability, are then applied to narrow down the range of remaining coefficients.
As the final step, dynamical system analysis is performed to fine-tune the coefficients to yield fixed-point behavior that is consistent with data from benchmark flows. These benchmark flows are selected to encompass the physical features likely to be encountered in practical applications. This holistic approach aims to ensure overall reasonable behavior in practical flows similar to benchmark flows. However, in complex, unseen flows, especially those with large-scale coherent structures, the accuracy of the models cannot be guaranteed.

%The traditional approaches place higher emphasis on theoretical considerations and use data as a final resort in closure model coefficient calibration. First, the behavior of turbulence at various extremes - rapid distortion limit, equilibrium state of homogeneous turbulence, decaying turbulence and near-wall asymptotics - is used to determine/limit some of the unknown coefficient values. Kinematic constraints such as realizability  are then used to narrow down the range of the remaining coefficients. 
%Finally, dynamical system analysis is employed to ensure that the coefficients yield fixed-point behavior consistent with data  in benchmark flows.  The benchmark flows are chosen to encompass the physical features likely to be encountered in practical applications. This holistic approach ensures overall reasonable behavior in practical flows similar to benchmark flows. In complex unseen flows, especially those with largescale coherent structures, the accuracy of the models cannot be guaranteed. 

%The closure model development sequence next invokes turbulence theories at various asymptotic limits to evaluate the  closure coefficients must Invocation of various physical principles and mathematical constraints to reduce the number of free coefficients - rapid distortion theory, log-law near wall behavior, fixed-point analysis, realizability etc.

\paragraph{\bf ML Approach Step 3: Coefficient determination.}
In its purest form, machine learning-based turbulence modeling aims to extract key flow physics directly from data, bypassing the need for foundational closure analysis and asymptotic theories. Given the intricacy of turbulence and the vast parameter space of the phenomenon, an immense volume of data is needed to comprehensively encompass all potential turbulence states. It is uncertain whether, in the foreseeable future, even a fraction of the requisite data can be effectively obtained through experiments or high-fidelity simulations. Due to the inherent incompleteness of data, there is a question as to whether ML methods can capture and incorporate essential physical aspects of turbulence into the closure models. Thus, the performance of ML-based turbulence models in complex unseen flows remains uncertain. Indeed, lacking the foundational underpinnings of traditional models, the performance of ML models in unseen flows may be inferior to that of traditional models.

%In its purest form, machine learning-based turbulence modeling aims to extract key flow physics directly from data, circumventing the need for foundational closure analysis and asymptotic theories. Given the intricacy of turbulence and the vast parameter space of the phenomenon,  immense volume of data is needed to comprehensively encompass all potential turbulence states. It is unclear   in the foreseeable future, if even a fraction of requisite data can be effectively obtained through experiments or high-fidelity simulations. Due to the inherent incompleteness of data, it is questionable if ML methods methods can capture and incorporate quintessential  physical aspects of turbulence into the closure models. Thus, the performance of ML-based turbulence models in complex unseen flows remains doubtful. Indeed, lacking the foundational underpinnings of traditional models, the performance of ML models in unseen flows can be inferior to that of traditional models. 

Current efforts in machine learning (ML)-enhanced turbulence closure modeling primarily focus on two main directions: (i) advancing Reynolds-averaged Navier-Stokes (RANS) through supervised learning, and (ii) employing supervised learning for scale-resolving simulation (SRS) approaches. In the case of RANS methods, there is a growing trend toward integrating crucial elements of traditional closure components with supervised learning techniques. However, finding an optimal balance between data-driven and conventional methods and effectively implementing them remains an ongoing challenge. Even if this balance is achieved and certain non-local effects are incorporated into RANS models, the generalizability of the models to unseen flows cannot be reliably guaranteed. Therefore, RANS calculations for complex unseen flows may still produce inaccurate results. On the other hand, ML-enhanced SRS approaches show promise for the future, but their development has not been as extensive as RANS. The primary obstacle in this regard is the difficulty in accurately identifying and quantifying the essential features that must be resolved within the simulations. In the longer term, the most effective application of ML approaches may involve developing scale-dependent closures for capturing the universal near-equilibrium characteristics of turbulence.

\vspace{0.2in}

{\bf Parting thoughts.}
Successful turbulence closure modeling involves four key elements - (i) Keen understanding of the key physical and mathematical attributes of turbulence; (ii) Prudent leveraging of turbulence knowledge assimilated over the last several decades; (iii) Engaging in abstract thinking to develop closure modeling principles that can lead to some degree of generalizability; and (iv) Employing effective and innovative tools to incorporate the elements identified above into closure models. 
Currently, neither traditional nor ML-based approaches accomplish all of the above adequately. The optimal strategy, for now, involves combining the strengths of both approaches optimally with human-expert input. Looking to the distant future, it is possible that artificial general intelligence (AGI) or artificial super intelligence (ASI) may accomplish all tasks without human intervention. As we progress in that direction, many of the insights and inferences expressed here must be revisited and re-evaluated.

\bibliographystyle{plain}
\bibliography{ref}
\end{document}